\documentclass[12pt]{revtex4-1}
\usepackage{amsmath}
\usepackage{hyperref}
\usepackage{braket}
\usepackage{dcolumn}
\usepackage{graphicx}
\usepackage{moreverb}
\usepackage{float}
\usepackage{subfigure}
\begin{document}

\title{Variational calculations for the hydrogen--antihydrogen system with a mass--scaled 
Born--Oppenheimer potential}
\author{Henrik Stegeby, Konrad Piszczatowski, Hans O Karlsson, Roland Lindh and Piotr Froelich}
\email{piotr.froelich@kvac.uu.se}  
\affiliation{Department of Chemistry - {\AA}ngstr\"om, The Theoretical Chemistry Programme, Uppsala University, \\ Box 518, 75120 Uppsala, Sweden}
\date{\today}
\begin{abstract}
The problem of proton-antiproton motion in the ${\rm H}$--${\rm \bar{H}}$ system is investigated by means of the variational method. 
We introduce a modified   nuclear interaction  through mass-scaling of the Born-Oppenheimer potential. This improved treatment of the interaction includes the nondivergent part of the otherwise divergent adiabatic correction  and shows the correct threshold behaviour.    
 Using this potential we calculate the vibrational energy levels with angular momentum $0$ and $1$ and the corresponding nuclear wavefunctions, as well as the S-wave scattering length. We obtain a full set of all bound states together with a large number of discretized continuum states that might be utilized in variational four-body calculations. The results of our calculations gives an indication of resonance states in the hydrogen-antihydrogen system.
\end{abstract}
\keywords{antihydrogen, Born-Oppenheimer approximation, adiabatic approximation, matter-antimatter interactions}
\pacs{36.10-k}
\maketitle
\section{Introduction}
\noindent
The interest in antihydrogen and its interaction with ordinary matter is inspired by the ongoing experiments on antihydrogen synthesis and trapping at CERN. The aim of these experiments is to use cold antihydrogen atoms for tests of the fundamental laws and symmetries of Physics. 

Substantial progress has been made during the recent years in that antihydrogen atoms have been trapped for 1 000 s \cite{alph11}. This  time is sufficiently long for the  ${\rm \bar{H}}$ atoms to spontaneously de-excite  from the highly excited Rydberg states, in which they are formed, to the ground state that is preferred in the planned spectroscopic and ballistic experiments.

\begin{figure}[H]
\center
\subfigure[]{
\includegraphics[height=3.5cm,width=3.5cm]{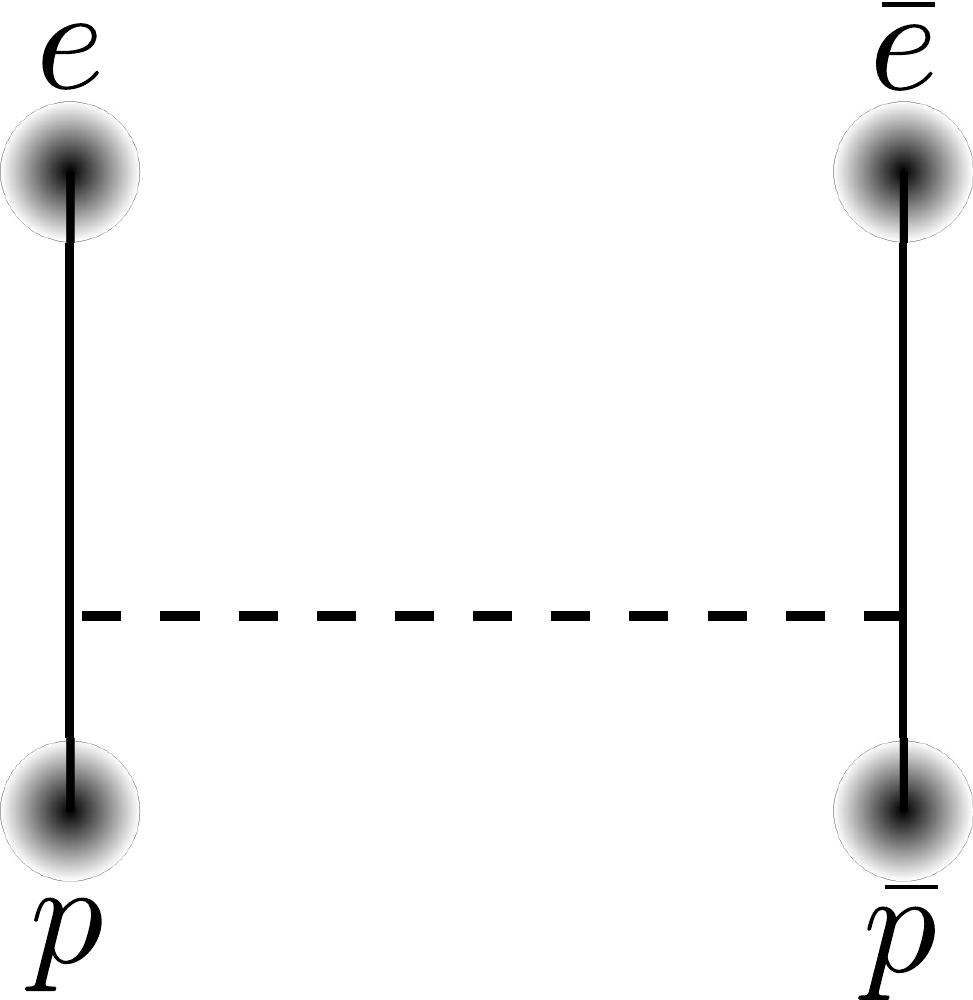}
}
\qquad
\subfigure[]{
\includegraphics[height=3.5cm,width=3.5cm]{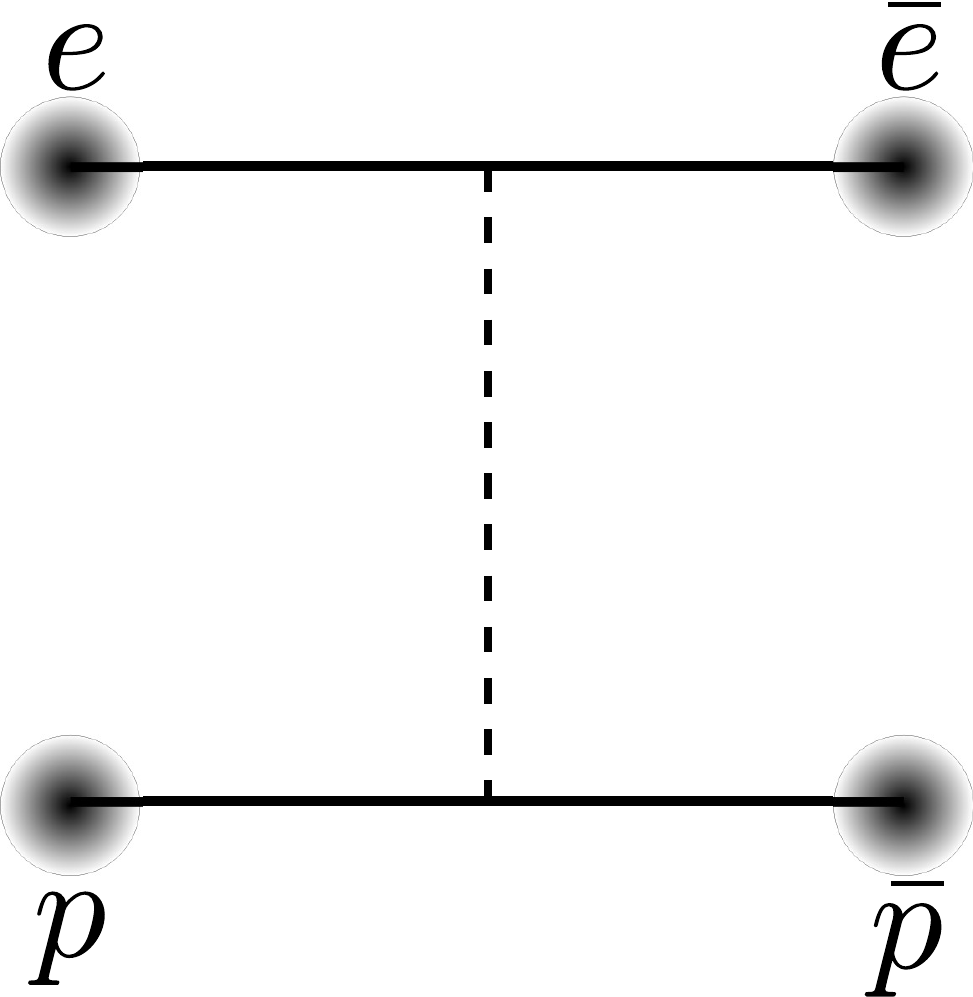}
}
\caption{Two possible configurations of the system. (a) Hydrogen-antihydrogen (b) Protonium-positronium. \label{channels}}
\end{figure}

The  generic example of matter-antimatter interaction is the collision between the two simplest atoms of each sort, the ${\rm H}$--${\rm \bar{H}}$  collision.  
In spite of its apparent simplicity the ${\rm H}$--${\rm \bar{H}}$  system proves to be very challenging. It significantly differs from e.g. the H--H system by the presence of annihilation between particles and antiparticles and by the rearrangement: the impinging ${\rm \bar{H}}$ and H atoms can recombine to two completely different atoms, Pn and Ps (see Figure \ref{channels}). The latter circumstance makes the problem very demanding both formally and computationally, even at the level of  Coulombic description that is addressed in the present work. 

The previous  studies of the ${\rm H}$--${\rm \bar{H}}$ system 
include distorted-wave approximation \cite{Froelich:00,Jonsell:01,Froelich:04a}, simple extensions  
thereof based on the close-coupling method \cite{Sinha:00}, the optical-potential method \cite{Zygelman:04}, 
and the Kohn variational method \cite{Armour:02}. All previous treatments have  been  based, one way or another, on the Born-Oppenheimer (BO)  approach. The extension to the adiabatic treatment 
was not possible since it was shown that the adiabatic correction diverges \cite{Strasburger:04}. 

The eigenvalues and eigenfunctions of the nuclear motion in the BO potential have so far  been  calculated only for a few of the highest exited states near the ${\rm H}$--${\rm \bar{H}}$ dissociation  threshold 
\cite{Froelich:00,Zygelman:01,Berggren:08,Voronin:08}, using numerical integration. 
In this paper we use a variational approach using Gaussian expansions to investigate the nuclear motion of the proton and antiproton in the improved BO potential. By using the variational method we obtain  a  set of  all  bound states (eigenvalues 
and corresponding  eigenfunctions)  in the improved leptonic potential that includes  the non-divergent 
part of the adiabatic correction and shows the correct asymptotic threshold behaviour. 


Previous calculations indicated  the presence  of possible near-threshold resonance states of the hydrogen-antihydrogen system \cite{Froelich:00,Zygelman:04,Voronin:08}. Such states  greatly influence the cross sections of both  elastic and inelastic  hydrogen-antihydrogen scattering \cite{Voronin:08}, acting as transient states 
mediating the rearrangement to Protonium and Positronium. We therefore devote  special care to  the treatment of the near threshold states since it is  {\em a priori}  possible that the improvement 
of the BO  potential may change the number of states below the threshold and/or their binding energies. 
The number of variationally obtained bound states below the threshold is double-checked using the procedure of Friedrich and Raab \cite{Friedrich:08, Raab:08}.

The BO-approach to the ${\rm H}$--${\rm \bar{H}}$ system works very well at the large ($\mu$m - nm) and intermediate 
(nm - $a_0$) internuclear distances. However 
as the atoms come closer, the leptonic clouds start to overlap, and the ${\rm H}$--${\rm \bar{H}}$ system gets prone to undergo a rearrangement into two completely different atoms,  Protonium (Pn) and Positronium (Ps). The rearrangement is 
particularly probable to occur below the so called critical distance ($R_c = 0.7427$ \cite{Armour:98,Strasburger:04}) 
below which the proton-antiproton dipole is not able to bind the two leptons, and  positronium 
can be released. The region around the critical distance is particularly difficult to treat.  
This is manifested by the divergence of the adiabatic correction to the Born-Oppenheimer interaction potential 
\cite{Strasburger:04} and highlights the need of a full four-body treatment. 

The four-body treatment might  be  e.g. based  on the variational method provided that the  latter would be  able to include the relevant arrangement channels and cope with their coupling.  If the variational method utilizing an expansion in a basis is applied, the basis functions should preferably allow easy transformations between the various arrangement channels, as e.g.  in the Gaussian Expansion Method (GEM) constructed by the Kamimura group for the four body problems \cite{Hiyama:03}. Having these aspects in mind, we expand the improved BO  solutions in a Gaussian basis, as to allow for the use of these solutions  
as a subset of the basis  in variational 4-body calculations. 
We deliver full set of states that are bounded by the BO  potential, together with a large number of  discretized continuum states (all expanded in a Gaussian basis) that might be utilized in variational four-body  applications. This might allow for a better understanding of the connection between the adiabatic and the full four body solutions, and provide the missing link in understanding why the adiabatic approach to ${\rm H}$--${\rm \bar{H}}$   breaks down  as manifested by the divergence of the adiabatic correction.

It should be recalled that the BO energy eigenvalues and eigenfunctions are normally obtained  in the BO potential that asymptotically converges to the ${\rm H}$--${\rm \bar{H}}$ dissociation threshold at $-1$ a.u.  (a value without adiabatic correction) and that was the case in all previous calculations.   However 
in a 4-body calculation the ${\rm H}$--${\rm \bar{H}}$ dissociation threshold, and all other thresholds, obviously takes  their  proper  value that include the adiabatic correction. In the present work we correct for this defect by including the non-divergent part of the adiabatic correction in our modified leptonic potential through a mass-scaling of the BO potential. It is important to emphasize that this procedure is {\it not}  equivalent to a simple  shift of the threshold 
for the  conventional BO potential (although it incidently assures the correct asymptotic  behaviour of the modified leptonic  potential). 

Atomic units have been used throughout this article.

\section{METHOD}
\subsection{Choice of the Hamiltonian}
\noindent
The 4-body Hamiltonian for hydrogen-antihydrogen system expressed in a space-fixed coordinate system reads
\begin{eqnarray}
H_{\rm 4body}^{\rm SF}=&-\frac{1}{2m_{\rm p}}\Delta_{{\bf r}_{\rm p}} 
-\frac{1}{2m_{\rm p}}\Delta_{{\bf r}_{\bar{\rm p}}}
 -\frac{1}{2}\Delta_{{\bf r}_{\rm e}} 
 -\frac{1}{2}\Delta_{{\bf r}_{\bar{\rm e}}} \nonumber \\&+ 
 V({\bf r}_{\rm p},{\bf r}_{\bar{\rm p}},{\bf r}_{\rm e},{\bf r}_{\bar{\rm e}})\,,
\end{eqnarray}
where ${\bf r}_{\rm p}$, ${\bf r}_{\bar{\rm p}}$, ${\bf r}_{\rm e}$ and ${\bf r}_{\bar{\rm e}}$ are position vectors  of proton, antiproton, electron and positron respectively,  $V$ describes Coulomb interactions between all particles, and $m_{\rm p}=1836.15267247$ a.u. is the proton mass.
Introducing body-fixed coordinates ${\bf r}_{\rm ep}={\bf r}_{\rm e}-{\bf r}_{\rm p}$, 
${\bf r}_{\rm {\bar e}{\bar p}}={\bf r}_{\bar{\rm e}}-{\bf r}_{\bar{\rm p}}$,
${\bf R}={\bf r}_{\rm p}-{\bf r}_{\bar{\rm p}}$ and separating the center-of-mass motion, the 4-body Hamiltonian can be rewritten as a sum of a leptonic Hamiltonian
\begin{equation}
H_{\rm lep}=-\frac{1}{2}\Delta_{{\bf r}_{\rm ep}}
-\frac{1}{2}\Delta_{{\bf r}_{\rm {\bar e}{\bar p}}}+V({\bf r}_{\rm ep},{\bf r}_{\rm {\bar e}{\bar p}},{\bf R})\,,
\label{Hlep}
\end{equation}
and Hamiltonian $H'$, which in this case has the following form
\begin{eqnarray}
H'=&-\frac{1}{2\mu_{\rm n}}\Delta_{\bf R} +\frac{1}{2\mu_{\rm n}}\nabla_{\bf R}\left( \nabla_{{\bf r}_{\rm ep}}-
\nabla_{{\bf r}_{\rm {\bar e}{\bar p}}}\right) \nonumber \\&
-\frac{1}{2m_{\rm p}}\Delta_{{\bf r}_{\rm ep}}-\frac{1}{2m_{\rm p}}\Delta_{{\bf r}_{\rm {\bar e}{\bar p}}}\,,
\label{Hp}
\end{eqnarray}
where $\mu_{\rm n}=m_{\rm p}/2$ is a nuclear reduced mass.

Assuming the total 4-body wavefunction as  a simple product of a given leptonic function
$\psi_{\rm lep}({\bf r}_{\rm ep},{\bf r}_{\rm {\bar e}{\bar p}};R)$, 
which is an eigenfunction of (\ref{Hlep})
\begin{equation}
H_{\rm lep}\psi_{\rm lep}({\bf r}_{\rm ep},{\bf r}_{\rm {\bar e}{\bar p}};R)=
E_{\rm BO}(R)\psi_{\rm lep}({\bf r}_{\rm ep},{\bf r}_{\rm {\bar e}{\bar p}};R)
\end{equation}
and some unknown nuclear function $\chi({\bf R})$, one obtains a nuclear Schr\"odinger equation within the adiabatic approximation
\begin{equation}
H_{\rm N}\chi_k({\bf R})=E_k\chi_k({\bf R})\,,
\end{equation}
where the nuclear Hamiltonian is defined as
\begin{equation}
H_{\rm N}=-\frac{1}{2\mu_{\rm n}}\Delta_{\bf R}+E_{\rm BO}(R)+\delta E_{\rm ad}(R)\,.
\label{HN}
\end{equation}
The last term on the r.h.s of  (\ref{HN}) is called the adiabatic correction, and is defined as an expectation value of the Hamiltonian $H'$ with respect to the leptonic wave function
\begin{equation}
\delta E_{\rm ad}(R)=\braket{\psi_{\rm lep}({\bf r}_{\rm ep},{\bf r}_{\rm {\bar e}{\bar p}};R)|
H'|\psi_{\rm lep}({\bf r}_{\rm ep},{\bf r}_{\rm {\bar e}{\bar p}};R)}\,.
\end{equation}

As it was shown by Strasburger \cite{Strasburger:04}, for the hydrogen--antihydrogen molecule the adiabatic correction diverges as the internuclear distance tends to the critical value 
$R_c\approx0.7427$ bohr. For this reason the nuclear Hamiltonian (\ref{HN}) is not well defined and one cannot calculate adiabatic energy levels for this system. Of course we can apply Born-Oppenheimer approximation, i.e. solve Schr\"odinger equation with the following nuclear Hamiltonian
\begin{equation}
H_{\rm N}^0=-\frac{1}{2\mu_{\rm n}}\Delta_{\bf R}+E_{\rm BO}(R)\,.
\label{HN0}
\end{equation}
However, one can still improve the Born-Oppenheimer potential by including only the non-divergent part of the adiabatic correction. This can be done ({\em vide} \ref{scalingapp}) through different factorization of the 4-body Hamiltonian and can be implemented by a simple scaling of the Born-Oppenheimer energy. 

The last two terms on the r.h.s of  (\ref{Hp}) do not depend on the internuclear coordinate ${\bf R}$, 
and we can treat them as a part of the kinetic energy of leptons.
This leads to a new leptonic Hamiltonian of the following form
\begin{equation}
\tilde{H}_{\rm lep}=-\frac{1}{2\mu}\Delta_{{\bf r}_{\rm ep}}
-\frac{1}{2\mu}\Delta_{{\bf r}_{\rm {\bar e}{\bar p}}}+V({\bf r}_{\rm ep},{\bf r}_{\rm {\bar e}{\bar p}},{\bf R})\,,
\label{Hlept}
\end{equation}
where $\mu=m_{\rm p}/(m_{\rm p}+1)$ is the electron--proton reduced mass. Solving the Schr\"odinger equation with Hamiltonian (\ref{Hlept}) one obtains a new leptonic energy curve $\tilde{E}_{\rm lep}(R)$, which on top of the Born-Oppenheimer contribution includes also a part of the adiabatic correction. It can be shown that the new energy is related to the original Born-Oppenheimer energy by the following mass scaling procedure
\begin{equation}
\tilde{E}_{\rm lep}(R)=\mu E_{\rm BO}(\mu R)\,.
\label{Elept}
\end{equation}
In can be also shown that the new leptonic potential given by (\ref{Elept}) has the correct adiabatic long range asymptotic behaviour, and nonadiabatic dissociation limit. For the detailed derivation and discussion of this procedure see \ref{scalingapp}. 

The scaling procedure allows us to define a new nuclear Hamiltonian with $E_{\rm BO}(R)+\delta E_{\rm ad}(R)$ substituted by
$\tilde{E}_{\rm lep}(R)$
\begin{equation}
\tilde{H}_{\rm N}=-\frac{1}{2\mu_{\rm n}}\Delta_{\bf R}+\tilde{E}_{\rm lep}(R)\,,
\label{HNt}
\end{equation}
which is well defined and can be used to calculate energy levels for the H--$\bar{\rm H}$ molecule. This approach leads to the energies which should be considered as being "halfway"  between the Born-Oppenheimer and the full adiabatic approximations. 
%
%
\subsection{The potential energy fit}
\begin{figure}[H]
\centering
\includegraphics[height=8.5cm,width=6cm,angle=270]{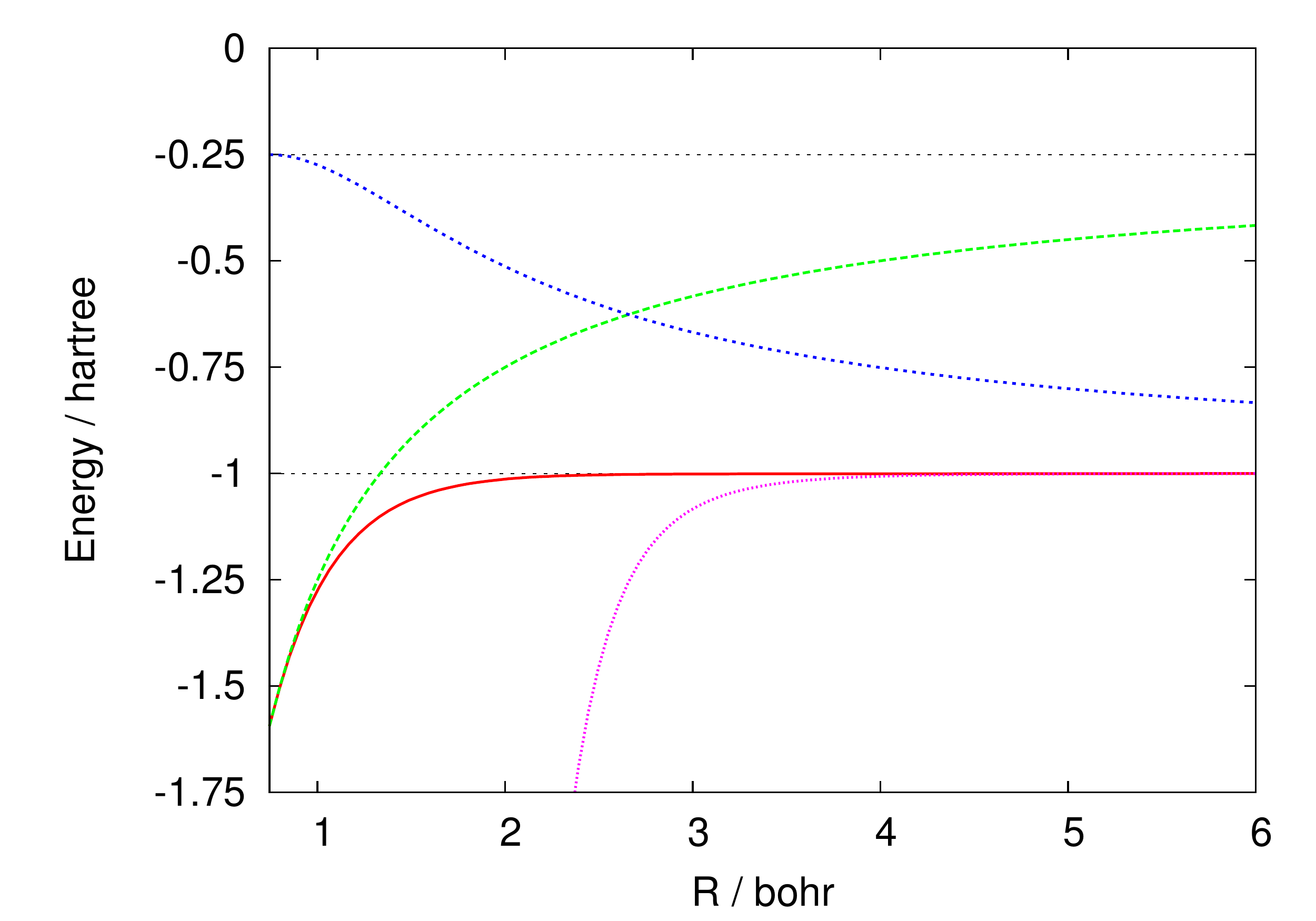}
\caption{The Born-Oppenheimer potential ($E_{\rm BO}(R)$) -- red, the short range approximation (\ref{EBOshort}) -- green, the long range approximation (\ref{EBOlong}) -- magenta, and the pure leptonic potential ($E_{\rm BO}(R)+1/R$) -- blue.\label{BOplot}}
\end{figure}

\noindent
In this work we used the Born-Oppenheimer potential for the H--$\bar{\rm H}$ system calculated by Strasburger \cite{Strasburger:02} with 256 explicitly correlated Gaussian (ECG) functions. The potential was computed for internuclear distances $R$ ranging from $R=0.744$ bohr up to 20.0 bohrs. 
For $R$ between 12 and 30 bohrs we have appended points computed by Strasburger with points obtained from asymptotic formula
\begin{equation}
E_{\rm BO}(R)=E_{\rm BO}^\infty-\sum_{n=6}^{26}\frac{C_n}{R^n}\,,
\label{EBOlong}
\end{equation}
where $E_{\rm BO}^\infty=-1$ hartree is the BO energy of two separated hydrogen  atoms and the Van der Waals constants $C_n$ were calculated by Mitroy and Ovsiannikov\cite{Mitroy:05}. 
\\
For $R=R_c$ the Born-Oppenheimer energy reaches value $E^{\rm Ps}_1-1/R_c$, where $E^{\rm Ps}_1=-0.25$ hartree is the positronium ground state energy. When the internuclear distance $R$ is smaller than the critical value $R_c$, the two leptons are no longer bound in the field of a dipole formed by the proton and the antiproton. Therefore, for $R<R_c$ we assumed that the Born-Oppenheimer energy is equal to the sum of $E_1^{\rm Ps}$ and the Coulomb attraction between the nuclei
 \begin{equation}
E_{\rm BO}(R)=E_1^{\rm Ps}-\frac{1}{R}\,. 
\label{EBOshort}
 \end{equation}
The choice of the potential for $R<R_c$ is in the spirit of the orthodox BO-approximation, i.e. the potential is equal to the lowest leptonic energy at each $R$, which in this region corresponds to the positronium ground state energy.
The calculated Born-Oppenheimer energy as well as the long and short range approximations to it are shown in Figure \ref{BOplot}.

\begin{table*}
\caption{\label{fittable}The fit parameters for Born-Oppenheimer ($V_{\rm BO}$) and mass-scaled leptonic ($\tilde{V}_{\rm lep}$) potentials.}
{\scriptsize
\begin{tabular}{l D{.}{.}{14} D{.}{.}{14} p{5mm} l D{.}{.}{14} D{.}{.}{14}}
\hline
 & \multicolumn{1}{c}{$V_{\rm BO}$} & \multicolumn{1}{c}{$\tilde{V}_{\rm lep}$} &&
 &  \multicolumn{1}{c}{$V_{\rm BO}$} & \multicolumn{1}{c}{$\tilde{V}_{\rm lep}$} \\
\hline
\hline
$A_{10}$   &  -19.8582635505679    &     -20.1369672678805   & &      $A_{40}$   &  -19.9771658686913    &     -20.2530003255519 \\
$A_{11}$   &   67.6269717956708    &      44.1781330016383   & &      $A_{41}$   & -173.8431212019852    &    -131.5359580062383 \\
$A_{12}$   &  -20.0575886039098    &     -14.6255427563730   & &      $A_{42}$   &   39.0038819732993    &      36.6202418042846 \\
$A_{13}$   &    1.6436298797648    &       1.2854178363453   & &      $A_{43}$   &   -4.9839482825694    &       2.3956839062761 \\
$A_{14}$   &   -0.0417677179701    &      -0.0344425997356   & &      $A_{44}$   &    0.2098970274357    &       0.0726049211608 \\
$A_{20}$   &   57.5162155781683    &      57.6405552715681   & &      $A_{50}$   &  -22.3850547348492    &     -22.6615482631176 \\
$A_{21}$   &    9.3918281802097    &     -24.8917421281046   & &      $A_{51}$   &  106.8813949154074    &     123.3264381665587 \\
$A_{22}$   &    3.0569545228764    &       8.8916405222450   & &      $A_{52}$   &  -24.8069885152175    &     -29.3520833084752 \\
$A_{23}$   &   -0.2521821480278    &      -1.4152945078081   & &      $A_{53}$   &    2.2667291349955    &      -4.7119434834608 \\
$A_{24}$   &    0.0965988366924    &       0.1428155629355   & &      $A_{54}$   &   -0.0258580270297    &       0.4403103436776 \\
$A_{30}$   &    4.7043278292101    &       5.4110168738119   & &      $A_{61}$   &    0.0000097266439    &       0.0000091571743 \\
$A_{31}$   &  -16.1993647293737    &     -17.2109690438145   & &      $A_{62}$   &   -0.0000006275304    &      -0.0000005856539 \\
$A_{32}$   &   23.4284275566323    &      24.7666406025529   & &      $A_{63}$   &    0.0000000184890    &       0.0000000171095 \\
$A_{33}$   &  -14.9456597423665    &     -15.7590990169304   & &      $A_{64}$   &   -0.0000000002113    &      -0.0000000001939 \\
$A_{34}$   &    3.8997649015388    &       4.1982517062010   & &      $\beta$    &    6.1520725018366    &       6.1431639772293 \\
$\alpha_1$ &    0.0897852714851    &       0.0893701431156   & &      $\alpha_4$ &    0.1412060702801    &       0.1171840549361 \\
$\alpha_2$ &    0.2268196733512    &       0.2952163755619   & &      $\alpha_5$ &    0.1048123413141    &       0.1108374703554 \\
$\alpha_3$ &    2.2437975957692    &       2.2164844767807   & &      $\alpha_6$ &    0.0068068098389    &       0.0067006105329 \\
\hline
\end{tabular}
}
\end{table*}
\noindent
For our purposes we have prepared an analytical fit  with 30 linear and 7 nonlinear parameters carefully optimized to reproduce the Born-Oppenheimer energy in all three discussed regions. The fitting function has the following form
\begin{eqnarray}
V_{\rm BO}(R)=&E_{\rm BO}^{\infty}+\left( E_1^{\rm Ps}-E_{\rm BO}^\infty-\frac{1}{R}\right)\exp(-\beta R^2) \nonumber \\&+
\sum_{n=1}^6\sum_{k=0}^4 A_{nk}R^k\exp(-\alpha_nR^2)\,.
\label{fitEBO}
\end{eqnarray}
with an additional constrain
\begin{equation}
\sum_{n=1}^6A_{n0}=0\,.
\end{equation}
The first term on the r.h.s. in (\ref{fitEBO}) describes the Born-Oppenheimer dissociation limit, the second one is added to ensure proper behaviour of the fit for small internuclear separations ({\em vide} (\ref{EBOshort})).

We have also prepared an analytical fit for the mass-scaled leptonic potential (\ref{Elept}). In this case the long range asymptotic expression for $\tilde{E}_{\rm lep}(R)$ has the same form as in (\ref{EBOlong}), but with $C_n$ constants replaced with $\tilde{C}_n=C_n/\mu^{n-1}$ and changed dissociation limit $\tilde{E}_{\rm lep}^\infty$. On the other hand the short range behaviour of $\tilde{E}_{\rm lep}(R)$ is derived from eqs. (\ref{EBOshort}) and (\ref{Elept}) , so the fitting function in this case has the following form
\begin{eqnarray}
\tilde{V}_{\rm lep}(R) =&\tilde{E}_{\rm lep}^{\infty}+\left( \mu E_1^{\rm Ps}-\tilde{E}_{\rm lep}^\infty-\frac{1}{R}\right)\exp(-\tilde{\beta} R^2) \nonumber \\&+
\sum_{n=1}^6\sum_{k=0}^4 \tilde{A}_{nk}R^k\exp(-\tilde{\alpha}_nR^2)\,,
\label{fitElept}
\end{eqnarray}
where $\tilde{E}_{\rm lep}^\infty=-\mu$ is twice the nonadiabatic ground state energy of the hydrogen atom. 
Also in this case parameters $\tilde{A}_{n0}$ are restrained by the condition
\begin{equation}
\sum_{n=1}^6\tilde{A}_{n0}=0\,.
\end{equation}
It should be stressed that the scaling is done for all internuclear distances, thus no discontinuity arrises in the potential (even though the critical distance changes under scaling).
All the linear and nonlinear parameters in (\ref{fitElept}) were optimized independently of the corresponding parameters in (\ref{fitEBO}). 
Since the values of the Born-Oppenheimer potential calculated by Strasburger \cite{Strasburger:02} are given on a grid for chosen $R$: $(R,E_{\rm BO}(R))$, we have been fitting the mass-scaled potential to the points obtained as follows $(R/\mu, \tilde{E}_{\rm lep}(R/\mu)=\mu E_{\rm BO}(R))$.
\\
All parameters for (\ref{fitEBO}) and (\ref{fitElept}) are given in Table \ref{fittable}. In both cases the fit errors with respect to the interaction energy are smaller than 0.1\% for internuclear distances $R<8.0$ bohrs and still not larger than 0.3\% for $R$ up to 30.0 bohrs. 
Since functions (\ref{fitEBO}) and (\ref{fitElept}) behave like $\exp(-\alpha R^2)$ for large $R$, they are not able to properly describe the asymptotic behaviour of the potentials for arbitrarily large $R$. However, we chose to use this type of the fitting function because it allows us to perform analytical calculations of the matrix elements ({\em vide infra}), which would not be possible with fits explicitly including 
the Van der Waals expansion (\ref{EBOlong}).

\subsection{Basis functions}
\noindent
To solve the nuclear Schr\"odinger equation with Hamiltonian defined by (\ref{HN0}) or (\ref{HNt}) for a given angular momentum $l$ we represent the nuclear wavefunctions as
\begin{equation}
\chi_{klm}({\bf R})=Y_{lm}(\hat{\bf R})\phi_{kl}(R)\,,
\label{chi}
\end{equation}
where $Y_{lm}$ is a spherical harmonic function, $\hat{\bf R}$ denotes angular coordinates of vector ${\bf R}$, and $\phi_{kl}(R)$ is expressed in a basis set of analytical functions
\begin{equation}
\phi_{kl}(R)=\sum_i c_i^{kl}{g}_i(R)=\sum_{n=1}^{n_{\rm max}}\left( c^{kl}_{2n-1}g_{nl}^c(R)+c^{kl}_{2n}g_{nl}^s(R)\right)\,.
\label{expansion}
\end{equation}
Basis functions $g_{nl}^c(R)$ and $g_{nl}^s(R)$ have the following form
\begin{eqnarray}
g_{nl}^c(R)=N_{nl}^cR^l\exp(-\nu_n R^2)\cos(\alpha\nu_n R^2)\\
g_{nl}^s(R)=N_{nl}^sR^l\exp(-\nu_n R^2)\sin(\alpha\nu_n R^2)
\label{oscgauss}
\end{eqnarray}
where $\alpha=\pi/2$ and $N_{nl}^{c}$, $N_{nl}^s$ are  normalization constants.

The nonlinear parameters $\nu_n$ defining the basis functions are chosen to be given by a geometrical progression
\begin{equation}
\nu_n=\frac{1}{r_n^2}
\end{equation}
with
\begin{equation}
r_n=r_{\rm min}\left(\frac{r_{\rm max}}{r_{\rm min}}\right)^{(n-1)/(n_{\rm max}-1)}\,.
\end{equation}
The parameter $n_{\rm max}$ is the number of cosine and sine oscillating Gaussian functions as defined in (\ref{oscgauss}), used in the radial wavefunction expansion. 

Substituting (\ref{chi}) into a nuclear Schr\"odinger equation one obtains the equation for the radial function $\phi_{kl}(R)$
\begin{equation}
\left(-\frac{1}{2\mu_{\rm n}}\frac{\rm d^2}{{\rm d}R^2}-\frac{1}{\mu_{\rm n} R}\frac{\rm d}{{\rm d}R}+\frac{l(l+1)}{2\mu_{\rm n} R^2}+V(R)-E_{kl}\right)\phi_{kl}(R)=0\,,
\end{equation}
with $V(R)$ being the Born-Oppenheimer or the mass-scaled leptonic potential.

The linear expansion coefficients $c_n^{kl}$ defining the function $\phi_{kl}(R)$ are obtained variationally be solving the generalized eigenvalue problem
\begin{equation}
({\bf T}+{\bf V}){\bf c}_{kl}=E_{kl}{\bf S}{\bf c}_{kl}\,,
\end{equation}
where ${\bf c}_{kl}$ is a column vector of coefficients $c_i^{kl}$,  ${\bf T}$ is a kinetic energy operator matrix 
\begin{equation}
T_{ij}=\braket{g_i|-\frac{1}{2\mu_{\rm n}}\frac{\rm d^2}{{\rm d}R^2}-\frac{1}{\mu_{\rm n} R}\frac{\rm d}{{\rm d}R}+\frac{l(l+1)}{2\mu_{\rm n} R^2}
|g_{j}}\,,
\end{equation}
${\bf V}$ is a potential energy operator matrix calculated with $V(R)$ given by (\ref{fitEBO}) or (\ref{fitElept}) 
\begin{equation}
V_{ij}=\braket{g_i|V(R)|g_j}
\end{equation}
and ${\bf S}$ is an overlap matrix.

\begin{figure}
\center
\subfigure{
\includegraphics[height=4cm,width=8.5cm]{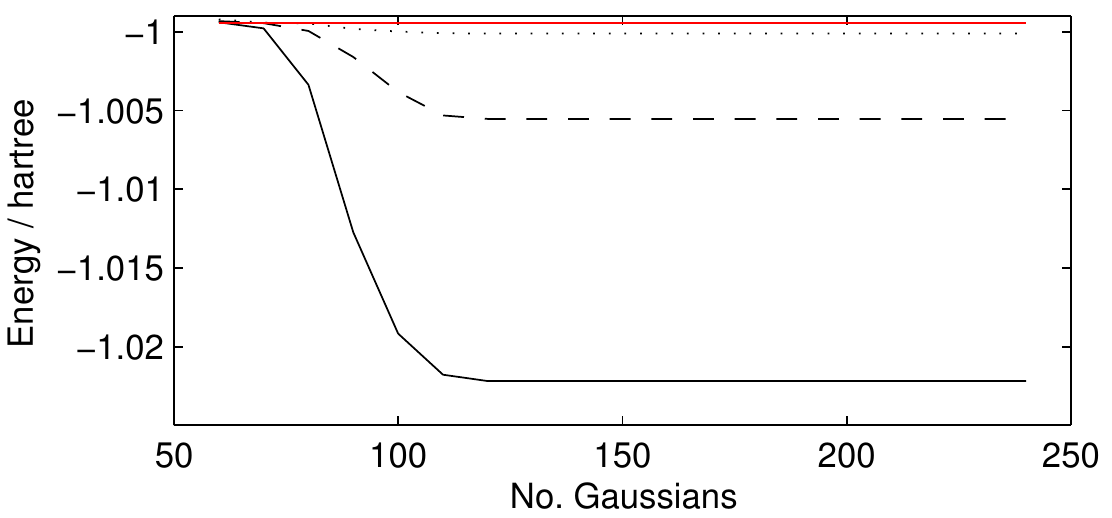}
}
\subfigure{
\includegraphics[height=4cm,width=8.5cm]{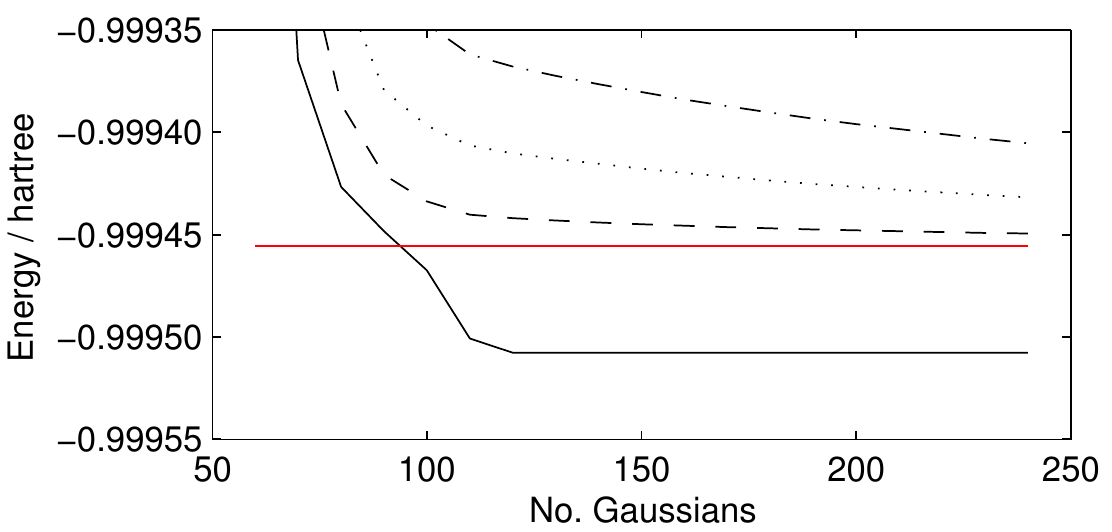}
}
\caption{Mass-scaled eigenvalues with $r_{{\rm min}}=0.00007$ a.u. and $r_{{\rm max}}=15$ a.u. kept constant. Top: $\nu=26-28$. Bottom: $\nu=29-32$. Threshold energy $E^{{\rm lep}}_{\infty} \approx -0.9994557$ (red line).}
\label{fig:eigtrends1}
\end{figure}

For a given expansion length one may optimize the basis functions by changing the values of $r_{\rm min}$ and $r_{\rm max}$. 
However, we must remember that different $r_{\rm min}$ and $r_{\rm max}$ would be optimal for different eigenstates. 
Because of the shape of the potential energy, the wavefunctions with the lowest energies are assumed to have much smaller average radius $\braket{R}$ than highly excited wavefunctions. For this reason we need to keep the parameter $r_{\rm min}$ small enough to be able to describe the ground state. At the same time the parameter $r_{\rm max}$ must be large if we want to be able to describe the wavefunction oscillations for highly excited states. This can be achieved only when the basis set expansion is long enough, which is illustrated in Figure \ref{fig:eigtrends1}.

\begin{figure}[H]
\centering
\includegraphics[height=4cm,width=8.5cm]{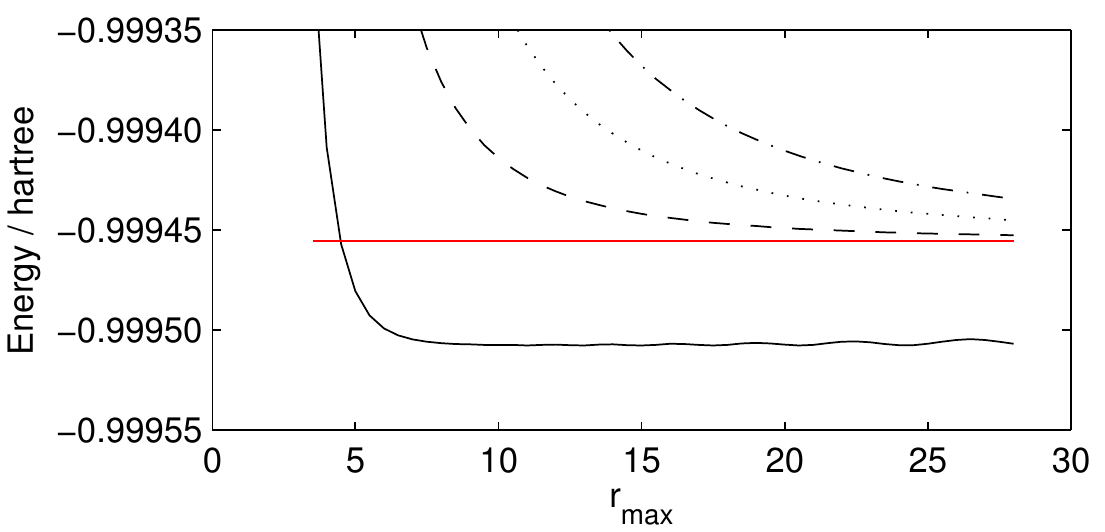}
\caption{The vibrational states $\nu=29-32$ dependence on $r_{\rm max}$, and the threshold energy (red line). $120$ Gaussians with $r_{{\rm min}}=0.00007$ a.u. have been used.}
\label{fig:eigtrends3}
\end{figure}

Figure \ref{fig:eigtrends3} shows how the energies for highly excited states depend on the $r_{\rm max}$ parameter for fixed values of $r_{\rm min}$ and given expansion length. If the value of $r_{\rm max}$ is too small the near threshold bound states are not described correctly. However, when $r_{\max}$ is too large one may observe oscillations of the bound states energies. We decided to use $r_{\rm max}=15$ for $n_{max}=60$ and $r_{\rm max}=20$ for $n_{max}=120$ in our calculations. These values are large enough to describe all bound states but still small enough to be in the region where the oscillations of the bound states energies are not severe.

\section{RESULTS}
\subsection{Bound states}
\begin{figure}
\center
\includegraphics[height=8.5cm,width=6cm,angle=270]{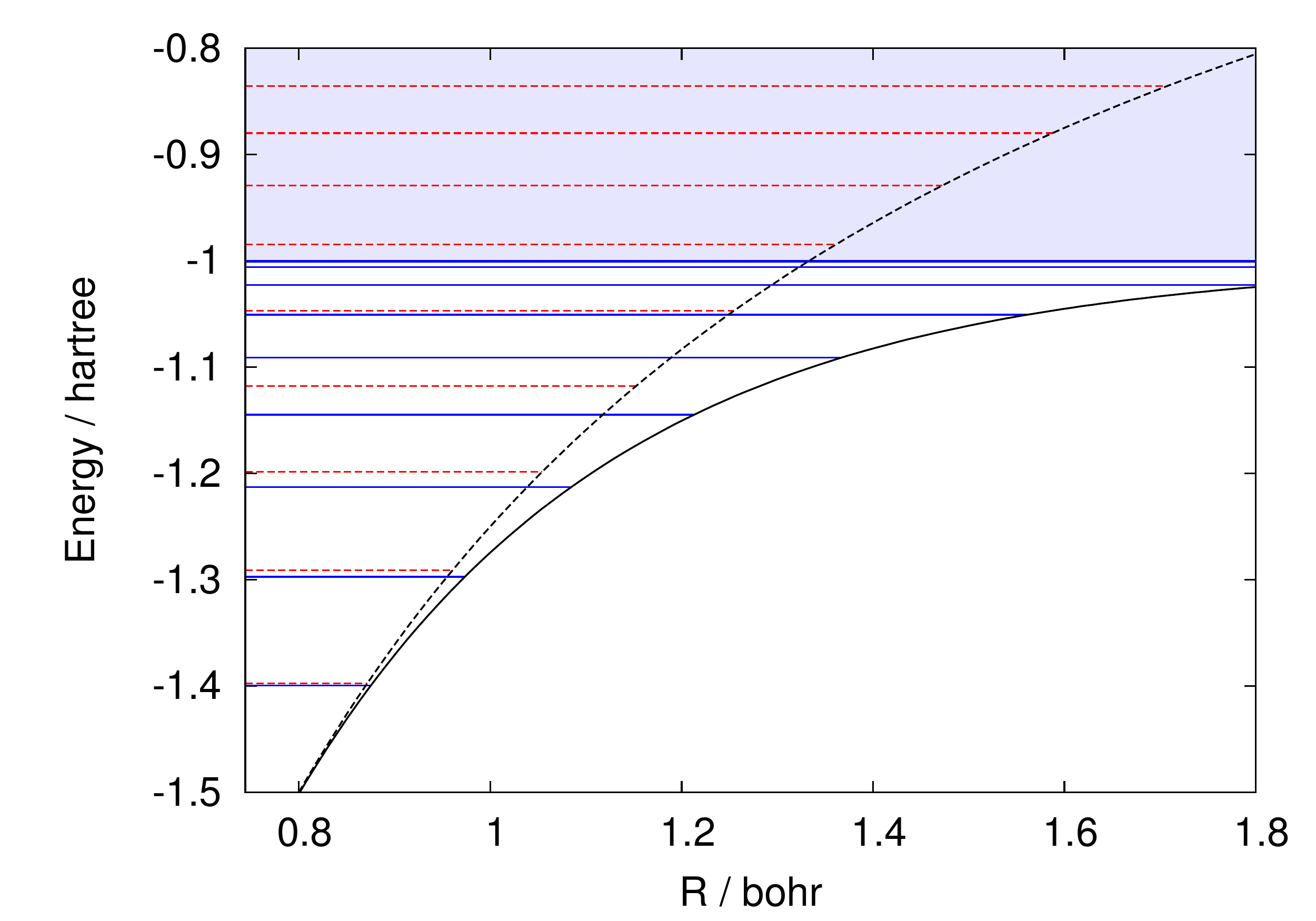}
\caption{The H--${\rm\bar H}$ oscillation levels for $l=0$  (solid blue lines) compared with protonium levels shifted by $E^{\rm Ps}_1$ (dashed red lines). The lowest shown levels correspond to quantum number $\nu=20$. Black lines represent potentials: the mass-scaled BO potential (solid line), the shifted Coulomb potential (\ref{EBOshort}) (dashed line) \label{figlev}.}
\end{figure}

\begin{figure}[H]
\centering
\includegraphics[height=4.5cm,width=8.5cm]{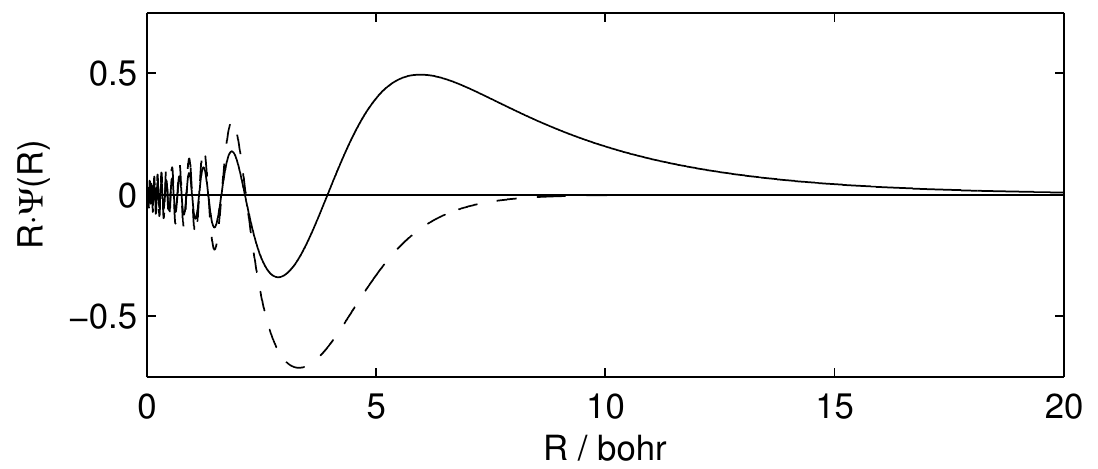}
\caption{Characteristics of the bound states $\nu=28$ (dashed) and $\nu=29$ (solid). $240$ Gaussians with $r_{\rm min}=0.00003$ a.u. and $r_{\rm max}=20$ a.u. have been used.}
\label{lastboundstates}
\end{figure}

\noindent
In Tables \ref{Levels0} and \ref{Levels1} we present the rovibrational energies of the hydrogen--antihydrogen molecule obtained with the Born-Oppenheimer ($E^{\rm H\bar{H}}_{\nu l}$) and the mass-scaled leptonic ($\tilde{E}^{\rm H\bar{H}}_{\nu l}$) potentials for angular momentum $l=0$ and $l=1$ respectively.
With both potentials we have found 29 bound states for $l=0$ and 28 bound states for $l=1$. Since for the small internuclear separations the potential is given by (\ref{EBOshort}), which is a shifted by the $E^{\rm Ps}_1$ potential for the protonium atom, we have expected the lowest eigenvalues to be equal to the protonium energy plus the positronium ground state energy 
\begin{equation}
E^{\rm H{\bar H}}_{\nu 0}=E^{\rm Pn}_\nu+E^{\rm Ps}_1\,. 
\label{Eshift}
\end{equation}
As it can be seen from Table \ref{Levels0} this relation is well fulfilled up to the vibrational quantum number $\nu=20$
({\em vide} the fifth column of Table \ref{Levels0}). 
For larger values of $\nu$ the difference between $E^{\rm H{\bar H}}_{\nu0}$ and $E^{\rm Pn}_\nu$ grows successively, which is caused by the fact that the nuclear wavefunction is no longer concentrated in the region where the condition (\ref{EBOshort}) is fulfilled. The differences between the Born-Oppenheimer energy levels for $l=0$ and the protonium $s$ states energies are shown in Figure \ref{figlev}.
It is obvious that the ground state radial wavefunction for the protonium atom and for the H--${\rm\bar H}$ system are nearly identical. However, the wavefunctions of near threshold states of the hydrogen--antihydrogen molecule differ from protonium orbitals. The radial wave functions for the last two bound sates of H--${\rm\bar H}$ with angular momentum $l=0$ are shown in Figure 
\ref{lastboundstates}.
\noindent
In the two last columns in Tables \ref{Levels0} and \ref{Levels1} we present the dissociation energies obtained with both our potentials for angular momentum $l=0$ and $l=1$ respectively. The dissociation energy for the Born-Oppenheimer approximation is defined as
\begin{equation}
\epsilon_{\nu l}=E^{\infty}_{\rm BO}-E^{\rm H\bar{H}}_{\nu l}\,,
\end{equation}
whereas for the mass-scaled potential we have
\begin{equation}
\tilde{\epsilon}_{\nu l}=\tilde{E}^\infty_{\rm lep}-\tilde{E}^{\rm H\bar{H}}_{\nu l}\,.
\end{equation}
The difference between $\epsilon_{\nu l}$ and $\tilde{\epsilon}_{\nu l}$ is negligible for the low lying states. However, the difference grows as the energies tends to the threshold energy and reaches 5\% for the last bound state with $l=0$ and 13\% for the last bound state with $l=1$.
\begin{table*}
\centering
\caption{The energy levels of the H--${\rm\bar{H}}$ molecule for angular momentum $l=0$: 
$E^{\rm H\bar{H}}_{\nu 0}$ -- obtained with the Born-Oppenheimer potential (second column); 
$\tilde{E}^{\rm H\bar{H}}_{\nu 0}$ -- obtained with the mass-scaled potential (third column).
$E^{\rm Pn}_\nu$ -- energy levels for the protonium atom (fourth column); 
$\delta_\nu=E^{\rm Pn}_\nu-E^{\rm H\bar{H}}_{\nu 0}$ -- Difference between protonium and H--${\rm\bar{H}}$ energies (fifth column);
$\epsilon_{\nu 0}$ -- dissociation energies obtained with  the Born-Oppenheimer potential (sixth column); 
$\tilde{\epsilon}_{\nu 0}$ dissociation energies obtained with the mass-scaled potential (seventh column).
\label{Levels0}}
{\scriptsize
\begin{tabular}{D{.}{}{1} D{.}{.}{10} D{.}{.}{10} D{.}{.}{10} D{.}{.}{4} p{2mm} D{.}{.}{10} D{.}{.}{10}}
\hline
\multicolumn{1}{c}{$\nu$} & 
\multicolumn{1}{c}{$E^{\rm H\bar{H}}_{\nu 0}$ }& 
\multicolumn{1}{c}{$\tilde{E}^{\rm H\bar{H}}_{\nu0}$} & 
\multicolumn{1}{c}{$E^{\rm Pn}_\nu$} & 
\multicolumn{1}{c}{$\delta_{\nu}$} && 
\multicolumn{1}{c}{$\epsilon_{\nu 0}$} & 
\multicolumn{1}{c}{$\tilde{\epsilon}_{\nu0}$}\\
\hline
\hline
  1. &  -459.28810584 &  -459.28797088 &  -459.03816812 & 0.250 && 458.28810584 & 458.28851520\\
  2. &  -115.00950131 &  -115.00936938 &  -114.75954203 & 0.250 && 114.00950131 & 114.00991370\\
  3. &   -51.25422639 &   -51.25409833 &   -51.00424090 & 0.250 && 50.25422639 & 50.25464265\\
  4. &   -28.93998669 &   -28.93986212 &   -28.68988551 & 0.250 && 27.93998669 & 27.94040644\\
  5. &   -18.61185049 &   -18.61172801 &   -18.36152672 & 0.250 && 17.61185049 & 17.61227233\\
  6. &   -13.00167067 &   -13.00154836 &   -12.75106023 & 0.251 && 12.00167067 & 12.00209268\\
  7. &    -9.61898924 &    -9.61886541 &    -9.36812588 & 0.251 && 8.61898924 & 8.61940973\\
  8. &    -7.42343727 &    -7.42331085 &    -7.17247138 & 0.251 && 6.42343727 & 6.42385517\\
  9. &    -5.91797886 &    -5.91784964 &    -5.66713788 & 0.251 && 4.91797886 & 4.91839396\\
 10. &    -4.84088720 &    -4.84075555 &    -4.59038168 & 0.251 && 3.84088720 & 3.84129987\\
 11. &    -4.04378975 &    -4.04365618 &    -3.79370387 & 0.250 && 3.04378975 & 3.04420050\\
 12. &    -3.43753603 &    -3.43740095 &    -3.18776506 & 0.250 && 2.43753603 & 2.43794527\\
 13. &    -2.96591308 &    -2.96577681 &    -2.71620218 & 0.250 && 1.96591308 & 1.96632113\\
 14. &    -2.59194272 &    -2.59180584 &    -2.34203147 & 0.250 && 1.59194272 & 1.59235016\\
 15. &    -2.29036921 &    -2.29023270 &    -2.04016964 & 0.250 && 1.29036921 & 1.29077702\\
 16. &    -2.04344058 &    -2.04330543 &    -1.79311784 & 0.250 && 1.04344058 & 1.04384975\\
 17. &    -1.83852260 &    -1.83838880 &    -1.58836736 & 0.250 && 0.83852260 & 0.83893312\\
 18. &    -1.66669147 &    -1.66655687 &    -1.41678447 & 0.250 && 0.66669147 & 0.66710119\\
 19. &    -1.52177984 &    -1.52163968 &    -1.27157387 & 0.250 && 0.52177984 & 0.52218400\\
 20. &    -1.39961222 &    -1.39945939 &    -1.14759542 & 0.252 && 0.39961222 & 0.40000371\\
 21. &    -1.29734882 &    -1.29717481 &    -1.04090288 & 0.256 && 0.29734882 & 0.29771913\\
 22. &    -1.21293458 &    -1.21273047 &    -0.94842597 & 0.265 && 0.21293458 & 0.21327479\\
 23. &    -1.14468025 &    -1.14443786 &    -0.86774701 & 0.277 && 0.14468025 & 0.14498218\\
 24. &    -1.09104749 &    -1.09075955 &    -0.79694126 & 0.294 && 0.09104749 & 0.09130387\\
 25. &    -1.05068830 &    -1.05034841 &    -0.73446107 & 0.316 && 0.05068830 & 0.05089273\\
 26. &    -1.02258176 &    -1.02218304 &    -0.67905054 & 0.344 && 0.02258176 & 0.02272736\\
 27. &    -1.00600201 &    -1.00553517 &    -0.62968199 & 0.376 && 0.00600201 & 0.00607949\\
 28. &    -1.00065727 &    -1.00012002 &    -0.58550787 & 0.415 && 0.00065727 & 0.00066434\\
 29. &    -1.00004946 &    -0.99950777 &    -0.54582422 & 0.454 && 0.00004946 & 0.00005209\\
 30. &    -0.99999650 &    -0.99945226 &    -0.51004241 & 0.490 && -0.00000350 & -0.00000342\\
\hline
\end{tabular}
}
\end{table*}

\begin{table*}
\caption{The H--${\rm\bar{H}}$ energy levels and dissociation energies for angular momentum $l=1$:  the energy levels $E^{\rm H\bar{H}}_{nu1}$ and $\tilde{E}^{\rm H\bar{H}}_{\nu1}$ obtained with the Born-Oppenheimer and mass-scaled leptonic potentials respectively; the dissociation energies $\epsilon_{\nu 1}$ and $\tilde{\epsilon}_{\nu 1}$ obtained  with the Born-Oppenheimer and mass-scaled leptonic potentials respectively. \label{Levels1}}
{\scriptsize
\begin{tabular}{r D{.}{.}{10} D{.}{.}{10} D{.}{.}{10} D{.}{.}{10}}
\hline
\multicolumn{1}{c}{$\nu$} & \multicolumn{1}{c}{$E^{\rm H\bar{H}}_{\nu1}$} &
 \multicolumn{1}{c}{$\tilde{E}^{\rm H\bar{H}}_{\nu1}$} & 
 \multicolumn{1}{c}{$\epsilon_{\nu1}$} & \multicolumn{1}{c}{$\tilde{\epsilon}_{\nu1}$} \\
\hline
\hline
  1. &  -115.00950985 &  -115.00937728 &   114.00950985 &   114.00992160\\
  2. &   -51.25422287 &   -51.25409434 &    50.25422287 &    50.25463866\\
  3. &   -28.93997500 &   -28.93985016 &    27.93997500 &    27.94039448\\
  4. &   -18.61183446 &   -18.61171187 &    17.61183446 &    17.61225619\\
  5. &   -13.00165510 &   -13.00153280 &    12.00165510 &    12.00207712\\
  6. &    -9.61897877 &    -9.61885498 &     8.61897877 &     8.61939931\\
  7. &    -7.42343471 &    -7.42330837 &     6.42343471 &     6.42385269\\
  8. &    -5.91798413 &    -5.91785500 &     4.91798413 &     4.91839932\\
  9. &    -4.84089719 &    -4.84076562 &     3.84089719 &     3.84130994\\
 10. &    -4.04379959 &    -4.04366610 &     3.04379959 &     3.04421042\\
 11. &    -3.43754139 &    -3.43740635 &     2.43754139 &     2.43795067\\
 12. &    -2.96591238 &    -2.96577614 &     1.96591238 &     1.96632046\\
 13. &    -2.59193806 &    -2.59180118 &     1.59193806 &     1.59234550\\
 14. &    -2.29036510 &    -2.29022859 &     1.29036510 &     1.29077291\\
 15. &    -2.04344083 &    -2.04330568 &     1.04344083 &     1.04385000\\
 16. &    -1.83852684 &    -1.83839303 &     0.83852684 &     0.83893735\\
 17. &    -1.66669349 &    -1.66655892 &     0.66669349 &     0.66710324\\
 18. &    -1.52176867 &    -1.52162862 &     0.52176867 &     0.52217294\\
 19. &    -1.39957504 &    -1.39942243 &     0.39957504 &     0.39996675\\
 20. &    -1.29727458 &    -1.29710087 &     0.29727458 &     0.29764520\\
 21. &    -1.21281710 &    -1.21261340 &     0.21281710 &     0.21315772\\
 22. &    -1.14452007 &    -1.14427809 &     0.14452007 &     0.14482241\\
 23. &    -1.09085090 &    -1.09056341 &     0.09085090 &     0.09110773\\
 24. &    -1.05046575 &    -1.05012623 &     0.05046575 &     0.05067055\\
 25. &    -1.02234924 &    -1.02195075 &     0.02234924 &     0.02249507\\
 26. &    -1.00579048 &    -1.00532349 &     0.00579048 &     0.00586781\\
 27. &    -1.00056446 &    -1.00002650 &     0.00056446 &     0.00057082\\
 28. &    -1.00001426 &    -0.99947180 &     0.00001426 &     0.00001612\\
 29. &    -0.99999527 &    -0.99945098 &    -0.00000473 &    -0.00000470\\
\hline
\end{tabular}
}
\end{table*}

\subsection{Scattering states and scattering length}
\begin{figure}[H]
\centering
\includegraphics[height=4.5cm,width=8.5cm]{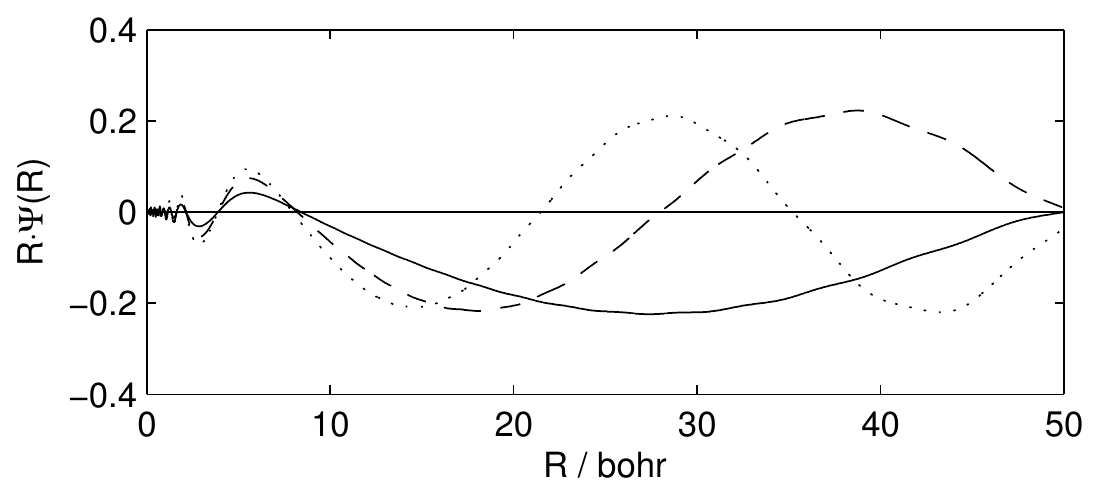}
\caption{The first few continuum states $\nu=30$ (dashed), $\nu=31$ (solid) and $\nu = 32$ (dotted). $240$ Gaussians with $r_{\rm min}=0.00003$ a.u. and $r_{\rm max}=20$ a.u. have been used.}
\label{firstcontinuum1}
\end{figure}

\begin{figure}[H]
\centering
\includegraphics[height=4.5cm,width=8.5cm]{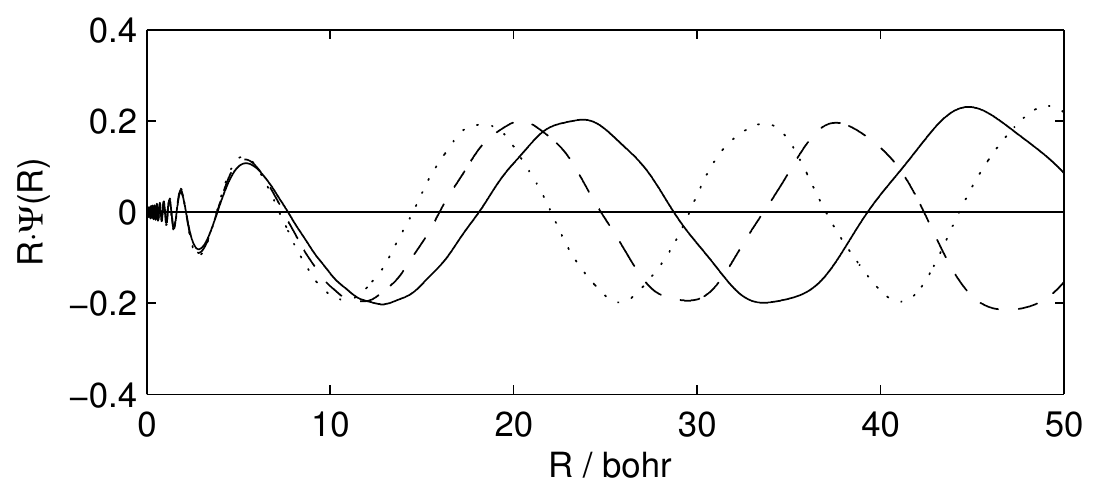}
\caption{The continuum states $\nu=33$ (solid), $\nu=34$ (dashed) and $\nu = 35$ (dotted). $240$ Gaussians with $r_{\rm min}=0.00003$ a.u. and $r_{\rm max}=20$ a.u. have been used.}
\label{firstcontinuum2}
\end{figure}

\begin{figure}[H]
\centering
\includegraphics[height=4.5cm,width=8.5cm]{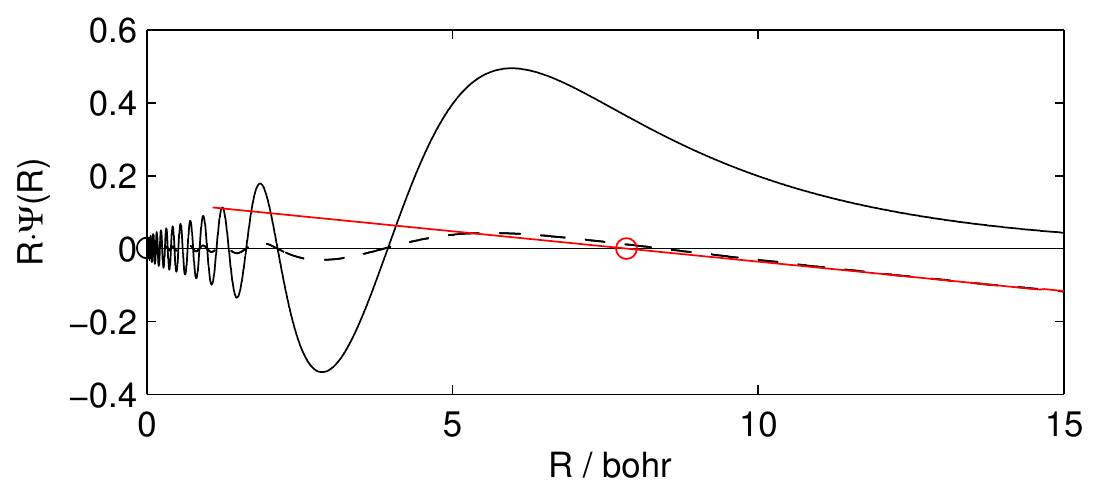}
\caption{The last bound state $\nu=29$ (solid), the first continuum state $\nu=30$ (dashed), and the tangent line to the first continuum state in the large $R$ asymptotic region (red). $240$ Gaussians with $r_{\rm min}=0.00003$ a.u. and $r_{\rm max}=20$ a.u. have been used, and the estimated scattering length is $R_{sc} \approx7.6$.}
\label{scattlength}
\end{figure}

\noindent
Since we are using an algebraic approximation to solve the nuclear Schr\"odinger equation, not only bound but also continuum states, however with discrete energies,  are obtained (see Figures \ref{firstcontinuum1} and \ref{firstcontinuum2}). 
Because we are using Gaussian basis functions in our description, the wave functions corresponding to the scattering states cannot possess the proper asymptotic behaviour for an arbitrarily large $R$. However, we were still able to provide a proper description not only for small $R$, where the potential well is deep and the wave functions vary rapidly, but also for $R$ large enough to see the asymptotic properties of the wavefunctions.
To test the quality of these scattering wavefunctions we used them to estimate the value of the scattering length. 
We  have used a geometrical procedure to estimate the scattering length, by interpreting the scattering length as an intersection of the tangent line to the ``zero energy continuum state" $R\phi_{E}(R)$ with the $R$-axis, where $\phi_{E}(R)$ is a  the  radial continuum wavefunction for $E \rightarrow 0$ \cite{Joachain:75}  (in our computation 
the  lowest discretized continuum  state, see Figure \ref{scattlength}). 

\noindent
The estimated scattering length obtained with this procedure is $a=7.6 \pm 0.4$ bohr, where the uncertainty comes from the sensitivity of the obtained results to the Gaussian parameters $r_{\rm max}$ and $n_{\rm max}$. This result can be compared with the value  
$a=7.7$ bohr, which was obtained by numerical integration of the Schr\"odinger equation\cite{Voronin:08}.
This proves that our calculations have obtained a reasonably good description of the scattering states wavefunctions.

\subsection{Relation to the WKB approach}
\noindent
The semi-classical approximation has been proved to give accurate predictions of the number of bound states as well as the value of the scattering length for diatomic molecules\cite{Friedrich:08, Raab:08}. We have decided to use the WKB approach to confirm the results we have obtained with the algebraic approximation. 

\begin{figure}[H]
\centering
\includegraphics[height=5cm,width=8.5cm]{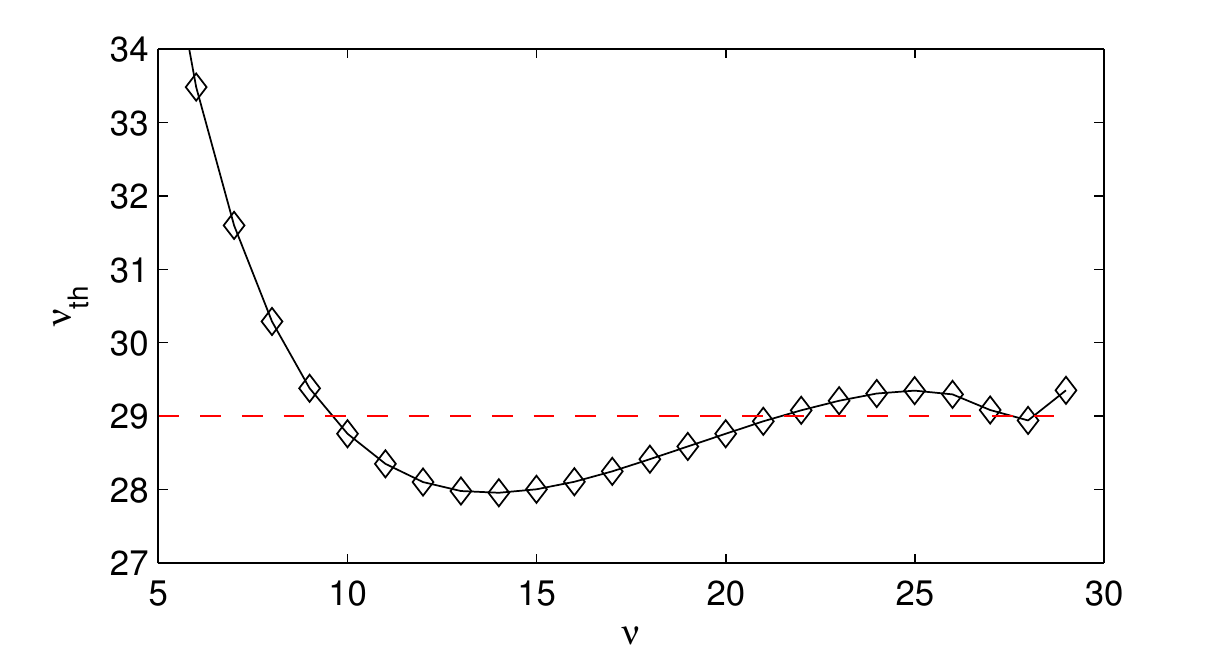}
\caption{The Quantization function for the BO potential for states with quantum number $\nu=6-29$. \label{Fig:nth}}
\label{quantization}
\end{figure}

When the potential energy has an attractive tail which decays faster then $1/R^2$, the system has a finite number of bound states where the dissociation energies $\epsilon_\nu$ and quantum numbers $\nu=1,2,3,\dots$ obey the following quantization rule
\begin{equation}
\nu_{\rm th}-\nu=F(\epsilon_\nu)\,,
\label{nth}
\end{equation}
where $\nu_{\rm th}$ is, in general non-integer, threshold quantum number, and $F$ is called the quantization function. As it was shown by Friedrich and Raab\cite{Friedrich:08, Raab:08} for a potential well with a homogeneous tail of order ($-6$), the $F$ function is given in the following form
\begin{eqnarray}
F(\epsilon_\nu)=&\frac{2b \kappa_\nu -(d \kappa_\nu)^2}{2\pi[1+(\kappa_\nu\beta_6)^4]} 
+ \frac{(\kappa_\nu\beta_6)^4}{1 + (\kappa_\nu\beta_6)^4} \nonumber \\
&\times \Big{[} -\frac{1}{8} + \frac{D}{2 \pi (\kappa_\nu \beta_6 )^{2/3}} 
+ \frac{ \Gamma (\frac{2}{3}) (\kappa_\nu \beta_6)^{2/3}}{4 \sqrt{\pi} \Gamma(\frac{7}{6})} \Big{]},
\label{QF}
\end{eqnarray}
where the constants $b$, $d$, $B$ and $D$ are defined in Table I in \cite{Friedrich:08}, $\kappa_\nu$ is defined as 
$\kappa_\nu=\sqrt{2\mathcal{M}\epsilon_\nu}$, where $\mathcal{M}$ is a nuclear reduced mass and $\beta_6$ is given in terms of the strength of the van der Waals interaction $C_6$ as
\begin{equation}
\beta_6 = \left(2C_6 \mathcal{M}\right)^{1/4} \,.
\end{equation}
The $\beta_6$ defines the potential range and is a typical scale for the quantum mechanical wavelengths and penetration depths.
\noindent
In our case $\beta_6=10.4521$ for the Born-Oppenheimer potential and $\beta_6=10.4592$ for the mass-scaled potential. 
We have calculated values of the $F(\epsilon_\nu)$ function for the dissociation energies listed in Table \ref{Levels0}. Substituting these results into formula (\ref{nth}) we obtained the threshold quantum numbers $\nu_{\rm th}$. The values of $\nu_{\rm th}$ computed from subsequent dissociation energies $\epsilon_{\nu}$ are shown in Table \ref{Tab:nth} and Figure \ref{Fig:nth}.
The predicted number of the bound states is the largest integer less than $\nu_{\rm th}$. 
 For lower quantum numbers $\nu$ the predicted number of bound states is rapidly changing, but for $\nu>22$ it stabilizes at a value equal 29, with one exception for $\nu=28$ where the $\nu_{\rm th}$ value is slightly below 29. The obtained results prove that we have succeeded to find all the bound states for H--$\bar{\rm H}$ molecule in its leptonic ground state and angular momentum zero.

Using the quantization function (\ref{QF}) and the relation
\begin{equation}
a = \bar{a}+\frac{b}{\tan(\pi F(\epsilon_\nu))}
\end{equation}
we were able to obtain an independent estimation of the scattering length. In the above formula $\bar{a}=b=0.4779888\beta_6$ (see Table I in \cite{Friedrich:08}). Substituting dissociation energies of the last bound state (zero angular momentum) calculated with the BO potential and with the mass-scaled BO potential, one gets the scattering length equals 7.6 bohr and 7.5 bohr respectively. These values are in full agreement with the results obtained from the algebraic approximation, which proves that we succeeded to correctly describe the asymptotic behaviour of the scattering wavefunctions in our calculations.

\begin{table}[H]
\centering
\caption{The threshold quantum number obtained for the BO potential ($\nu_{\rm th}$) and the mass-scaled BO potential ($\tilde{\nu}_{\rm th}$).\label{Tab:nth}}
{\scriptsize
\begin{tabular}{r p{5mm} D{.}{.}{2} p{5mm} D{.}{.}{2}}
\hline
\multicolumn{1}{c}{$\nu$} && \multicolumn{1}{c}{$\nu_{\rm th}$} && \multicolumn{1}{c}{$\tilde{\nu}_{\rm th}$}\\
\hline
\hline
20 && 28.75 && 28.76 \\
21 && 28.92 && 28.93 \\
22 && 29.07 && 29.08 \\
23 && 29.20 && 29.21 \\
24 && 29.30 && 29.31 \\
25 && 29.34 && 29.35 \\
26 && 29.29 && 29.29 \\
27 && 29.07 && 29.08 \\
28 && 28.94 && 28.94 \\
29 && 29.34 && 29.35 \\
\hline
\end{tabular}
}
\end{table}

\section{CONCLUSIONS}
By applying the algebraic approximation we have succeeded in obtaining the vibrational spectra of the leptonic ground state of the hydrogen-antihydrogen molecule in its two lowest rotational states. Since there is no possibility to include the adiabatic correction for this system, our calculations were done at the Born-Oppenheimer level. However, we were able to append the Born-Oppenheimer potential with the nondivergent part of the otherwise divergent adiabatic correction. This contribution is responsible for the change in the long range behaviour of the interaction potential, it  mimics the total adiabatic correction for large internuclear distances $R$ and converges to the correct dissociation threshold. 
The discussed part of the adiabatic correction was included {\it via}  the mass-scaling procedure, which does not require any additional calculations besides the Born-Oppenheimer one. The mass-scaling procedure leads to  the proper nonadiabatic dissociation threshold of  the leptonic potential.
As it was shown the near threshold  states are significantly affected by this procedure, since the changes in the dissociation energies reaches a few percents (as much as 13\% for the most loosely bound state of the $l=1$ symmetry).

Using the algebraic approximation we were able to obtain wavefunctions not only for all bound states, but for continuum states as well. The energies of the calculated scattering states are of course discretized. However, the corresponding wavefunctions reproduce the correct asymptotic behaviour up to a certain internuclear  distance $R^*$ that is larger than the range  of the spacially  extended  
near-threshold states.
This allowed us to calculate the scattering length for the hydrogen-antihydrogen collisions. 
The value obtained in this calculation is in very good agreement with the results of high precision numerical computations and with the results of semi-classical approach, which confirms good quality 
of the discretized continuum states.

The existence of the bound states for the H--$\bar{\rm H}$ system in the Born-Oppenheimer approximation may indicate the existence of quasi-stable states (resonances) for this system in the nonadiabatic description. To confirm or falsify these predictions one has to perform 4-body calculations for this system.
The wave functions obtained in our calculations can be used as a well adapted basis set for the internuclear degree of freedom in the  H--$\bar{\rm H}$ channel in the fully nonadiabatic (4-body) calculations for H--$\bar{\rm H}$. We suggest  that the nuclear wave functions obtained in the present work will provide  a better  description of the H--$\bar{\rm H}$ channel  in the 4-body calculations as compared to the   alternative description using  protonium basis set. 
Application of the here obtained nuclear wave functions as a part of the basis set in 
4-body calculations may not only be  more efficient 
but also allow for better understanding of the relation  between the B-O and fully non-adiabatic methods 
in the demanding case of H--$\bar{\rm H}$. 
Using a mass-scaled BO basis will not cure the nonadiabaticity, but it might facilitate its investigation. The fact that the BO functions are part of the full basis helps to identify the adiabatic components in the nonadiabatic solutions. In the 4-body calculations the mass-scaled BO basis must be augmented by other dedicated subspaces for description of other Jacobi fragments and coordinates.

\section*{Acknowledgement}
This work has been supported by the Wenner-Gren Foundation, 
the Swedish National Science Council and  the Sweden-Japan Foundation. We are grateful to prof. 
M. Kamimura for valuable discussions concerning the use of gaussian expansions in four-body calculations.

\appendix

\section{Mass-Scaling procedure of the Born-Oppenheimer potential\label{scalingapp}}
\noindent


The adiabatic correction to the BO potential is defined as an expectation value of the nuclear part of the Hamiltonian with respect to the leptonic wave function (i.e. the eigenfunction of the leptonic Hamiltonian). 
As it was shown by Strasburger \cite{Strasburger:04} in the case of hydrogen--antihydrogen interaction the adiabatic correction is not well defined for certain  internuclear separations, and the standard procedure of improving the BO potential cannot be applied. For this reason we decided to apply another approach, which enables us to include at least the part of the adiabatic correction which is responsible for long range behaviour of the interaction potential. This is done  following  a different definition of the leptonic Hamiltonian \cite{Kutzelnigg:97, Froman:62}  and results in a simple scaling of the Born-Oppenheimer potential.

The total Hamiltonian for the H--$\bar{\rm H}$ molecule expressed in a space-fixed coordinate system reads
\begin{equation}
H_{\rm tot} = T_{\rm lep} + T_{\rm nuc}+V\,,
\end{equation}
where
\begin{eqnarray}
T_{\rm lep} = -\frac12\Delta{{\bf r}_{\rm e}} -\frac12\Delta{{\bf r}_{\bar{\rm e}}}\,, \\
T_{\rm nuc} = -\frac{1}{2m_{\rm p}}\Delta_{{\bf r}_{\rm p}}-\frac{1}{2m_{\rm p}}\Delta_{{\bf r}_{\bar{\rm p}}}\,,
\end{eqnarray}
and $V$ describes Coulomb interactions between all the particles in the system.

The first step in most rigorous treatments is the separation of the COM motion. The COM coordinates are defined as
\begin{equation}
{\bf X}=-\frac{1}{M}\left( {\bf r}_{\rm e}+{\bf r}_{\bar{\rm e}}+m_{\rm p}{\bf r}_{\rm p}+m_{\rm p}{\bf r}_{\bar{\rm p}}\right)\,,
\end{equation}
where $M=2m_{\rm p}+2$ is the total mass of the system. With this separation made, the total Hamiltonian can be rewritten as a sum of the COM kinetic operator and the Hamiltonian for the internal motion of leptons and nuclei
\begin{equation}
H_{\rm tot} = -\frac{1}{2M}\Delta_{\bf X}+H_{\rm rel}\,.
\end{equation}
The total wave function separates as
\begin{equation}
\Psi_{\rm tot}({\bf r}_{\rm e},{\bf r}_{\bar{\rm e}},{\bf r}_{\rm p},{\bf r}_{\bar{\rm p}})=\psi_{\rm rel}({\bf r}_{\mu\nu},\dots)\chi({\bf X})\,,
\end{equation}
where ${\bf r}_{\mu\nu}$ are body-fixed coordinates.
Following the Born-Oppenheimer approximation, the Hamiltonian describing the relative motion is divided into $H_{\rm lep}$ and $H'$, where the leptonic Hamiltonian $H_{\rm lep}$ does not depend on the nuclear masses

\begin{equation}
H_{\rm rel} = H_{\rm lep} + H'\,.
\label{div1}
\end{equation}
If the internal coordinates are chosen to be
\begin{equation}
{\bf r}_{\rm ep} = {\bf r}_{\rm e}-{\bf r}_{\rm p}\,,\quad
{\bf r}_{\bar{\rm e}\bar{\rm p}} = {\bf r}_{\bar{\rm e}}-{\bf r}_{\bar{\rm p}}\,, \quad
{\bf R} = {\bf r}_{\rm p}-{\bf r}_{\bar{\rm p}}\,,
\end{equation}
the leptonic part reads
\begin{equation}
H_{\rm lep}=-\frac12\Delta_{{\bf r}_{\rm ep}}-\frac12\Delta_{{\bf r}_{\bar{\rm e}\bar{\rm p}}}+
V({\bf r}_{\rm ep},{\bf r}_{\bar{\rm e}\bar{\rm p}},{\bf R})
\label{lep1}
\end{equation}
and the $H'$ operator itself can be divided into three contributions
\begin{eqnarray}
 H'_{(a)}=-\frac{1}{m_{\rm p}}\Delta_{\bf R}\,,\\
 H'_{(b)}=
\frac{1}{m_{\rm p}}\nabla_{\bf R}\left(\nabla_{{\bf r}_{\rm ep}}-\nabla_{{\bf r}_{\bar{\rm e}\bar{\rm p}}}\right)\,, \\
 H'_{(c)}=
-\frac{1}{2m_{\rm p}}\Delta_{{\bf r}_{\rm ep}}-\frac{1}{2m_{\rm p}}\Delta_{{\bf r}_{\bar{\rm e}\bar{\rm p}}}\,.
\label{nuc}
\end{eqnarray}
Part ($a$) describes the relative motion of the nuclei, ($b$) couples the nuclear and leptonic motion, and ($c$) can be considered as a correction to the leptonic kinetic energy. In fact $H'_{(c)}$ has the same form as the kinetic energy of leptons, but it is divided by the proton mass.  

The factorization of $H_{rel}$ ( \ref{div1})  given by (\ref{lep1}) and (\ref{nuc}) is not unique, and we can choose to include the $H'_{(c)}$ contribution in the leptonic Hamiltonian
\begin{equation}
\tilde{H}_{\rm lep}=H_{\rm lep}+H'_{(c)} =
-\frac{1}{2\mu}\Delta_{{\bf r}_{\rm ep}}-\frac{1}{2\mu}\Delta_{{\bf r}_{\bar{\rm e}\bar{\rm p}}}+
V({\bf r}_{\rm ep},{\bf r}_{\bar{\rm e}\bar{\rm p}},{\bf R})\,.
\label{lep2}
\end{equation}
This leads to the operator of the same form as (\ref{lep1}), but with the electron mass replaced by the electron--proton reduced mass, $\mu=m_{\rm p}/(m_{\rm p}+1)$.

One can relate the old leptonic Hamiltonian (\ref{lep1}) to the new one (\ref{lep2}) by applying the following coordinates transformation
\begin{eqnarray}
 {\bf r}_{\rm ep}\longrightarrow {\bf s}_{\rm ep} = \mu{\bf r}_{\rm ep}\,,\nonumber \\
 {\bf r}_{\bar{\rm e}\bar{\rm p}}\longrightarrow{\bf s}_{\bar{\rm e}\bar{\rm p}} =
\mu{\bf r}_{\bar{\rm e}\bar{\rm p}}\,,\nonumber \\
 {\bf R}\longrightarrow{\bf S}=\mu{\bf R}\,.
\end{eqnarray}
With these definitions we can write
\begin{eqnarray}
H_{\rm lep}({\bf s}_{\rm ep}, {\bf s}_{\bar{\rm e}\bar{\rm p}}, {\bf S})&=&
-\frac12\Delta_{{\bf s}_{\rm ep}}-\frac12\Delta_{{\bf s}_{\bar{\rm e}\bar{\rm p}}}+
V({\bf s}_{\rm ep},{\bf s}_{\bar{\rm e}\bar{\rm p}},{\bf S}) \nonumber \\&=&
-\frac{1}{2\mu^2}\Delta_{{\bf r}_{\rm ep}}-\frac{1}{2\mu^2}\Delta_{{\bf r}_{\bar{\rm e}\bar{\rm p}}} +\frac{1}{\mu}V({\bf r}_{\rm ep},{\bf r}_{\bar{\rm e}\bar{\rm p}},{\bf R})\,,
\end{eqnarray}
where we made use of the fact that the potential energy operator $V({\bf r}_{\rm ep},{\bf r}_{\bar{\rm e}\bar{\rm p}},{\bf R})$ is, in all coordinates, a homogeneous function of order $-1$. This leads to the following relation
\begin{equation}
\tilde{H}_{\rm lep}({\bf r}_{\rm ep}, {\bf r}_{\bar{\rm e}\bar{\rm p}}, {\bf R})=
\mu H_{\rm lep}(\mu{\bf r}_{\rm ep}, \mu{\bf r}_{\bar{\rm e}\bar{\rm p}}, \mu{\bf R})\,.
\label{scaling1}
\end{equation}
It is therefore clear that the eigenvalues and eigenvectors of the new leptonic Hamiltonian (\ref{lep2})
\begin{equation}
\tilde{H}_{\rm lep}\tilde{\psi}_{\rm lep}({\bf r}_{\rm ep}, {\bf r}_{\bar{\rm e}\bar{\rm p}};  R)=
\tilde{E}_{\rm lep}(R)\tilde{\psi}_{\rm lep}({\bf r}_{\rm ep}, {\bf r}_{\bar{\rm e}\bar{\rm p}}; R)
\end{equation}
and the eigenvalues and eigenvectors of the original leptonic Hamiltonian (\ref{lep1})
\begin{equation}
H_{\rm lep}\psi_{\rm lep}({\bf r}_{\rm ep}, {\bf r}_{\bar{\rm e}\bar{\rm p}};  R)=
E_{\rm BO}(R)\psi_{\rm lep}({\bf r}_{\rm ep}, {\bf r}_{\bar{\rm e}\bar{\rm p}}; R)
\end{equation}
can be related by the following scaling transformations
\begin{equation}
\tilde{E}_{\rm lep}(R)=\mu E_{\rm BO}(\mu R)
\label{scaling2}
\end{equation}
and
\begin{equation}
\tilde{\psi}_{\rm lep}({\bf r}_{\rm ep}, {\bf r}_{\bar{\rm e}\bar{\rm p}}; R)=
\mu^3\psi_{\rm lep}(\mu{\bf r}_{\rm ep}, \mu{\bf r}_{\bar{\rm e}\bar{\rm p}}; \mu R)\,,
\label{scaling3}
\end{equation}
where $\mu^3$ on the r.h.s of the above equation is a normalization factor.

It is obvious that the mass--transformed leptonic energy (\ref{scaling2}) consist of the Born-Oppenheimer energy $E_{\rm BO}(R)$ and some non-Born-Oppenheimer contributions. To discuss this in a systematic way, let us express $\tilde{E}_{\rm lep}$ as an expectation value of the $\tilde{H}_{\rm lep}$ operator with the wavefunction $\tilde{\psi}_{\rm lep}$
\begin{equation}
\tilde{E}_{\rm lep}(R)=
\braket{\tilde{\psi}_{\rm lep}({\bf r}_{\rm ep}, {\bf r}_{\bar{\rm e}\bar{\rm p}};  R)| \tilde{H}_{\rm lep}|
\tilde{\psi}_{\rm lep}({\bf r}_{\rm ep}, {\bf r}_{\bar{\rm e}\bar{\rm p}};  R)}\,.
\end{equation}
Using (\ref{scaling3}) we get
\begin{equation}
\tilde{E}_{\rm lep}(R)=\mu^2\mathcal{T}(\mu R)+\mu\mathcal{V}(\mu R)+
\mu^2\mathcal{E}_{\rm ad}^{(c)}(\mu R)\,,
\label{Etilde}
\end{equation}
where $\mathcal{T}$ and $\mathcal{V}$ denotes the expectation values of the kinetic and potential energy operators respectively
\begin{eqnarray}
\mathcal{T}(R)=
\braket{
\psi_{\rm lep}({\bf r}_{\rm ep}, {\bf r}_{\bar{\rm e}\bar{\rm p}};  R)|
-\frac12\Delta_{{\bf r}_{\rm ep}}-\frac12\Delta_{{\bf r}_{\bar{\rm e}\bar{\rm p}}}|
\psi_{\rm lep}({\bf r}_{\rm ep}, {\bf r}_{\bar{\rm e}\bar{\rm p}};  R)},\\
\mathcal{V}(R)=
\braket{
\psi_{\rm lep}({\bf r}_{\rm ep}, {\bf r}_{\bar{\rm e}\bar{\rm p}};  R)|
V({\bf r}_{\rm ep},{\bf r}_{\bar{\rm e}\bar{\rm p}},{\bf R})|
\psi_{\rm lep}({\bf r}_{\rm ep}, {\bf r}_{\bar{\rm e}\bar{\rm p}};  R)},
\end{eqnarray}
and
\begin{equation}
\mathcal{E}_{\rm ad}^{(c)}(R)=
\braket{
\psi_{\rm lep}({\bf r}_{\rm ep}, {\bf r}_{\bar{\rm e}\bar{\rm p}};  R)|
H'_{(c)}|
\psi_{\rm lep}({\bf r}_{\rm ep}, {\bf r}_{\bar{\rm e}\bar{\rm p}};  R)}
\end{equation}
is a part of the total adiabatic correction $\mathcal{E}_{\rm ad}=\braket{\psi|H'|\psi}$.
By expanding (\ref{Etilde}) in powers of $m_{\rm p}^{-1}$ one obtains the following expression 
\begin{eqnarray}
\tilde{E}_{\rm lep}(R)=&E_{\rm BO}(R)+\mathcal{E}_{\rm ad}^{(c)}(R) \nonumber \\&
-\left[ 2\mathcal{T}(R)+\mathcal{V}(R)+ R\frac{{\rm d}E_{\rm BO}(R)}{{\rm d}R}\right]\frac{1}{m_{\rm p}}
+\mathcal{O}(\frac{1}{m_{\rm p}^2})\,.
\label{Eexpan1}
\end{eqnarray}
On the other hand, applying eqs. (\ref{scaling1}), (\ref{scaling2}), (\ref{scaling3}) and the Hellmann-Feynmann theorem we can write
\begin{eqnarray}
\langle
\psi_{\rm lep}(\mu{\bf r}_{\rm ep}, \mu{\bf r}_{\bar{\rm e}\bar{\rm p}}; \mu R)
|\frac{\rm d}{\rm d\mu}H_{\rm lep}(\mu{\bf r}_{\rm ep}, \mu{\bf r}_{\bar{\rm e}\bar{\rm p}}, \mu{\bf R})
|\psi_{\rm lep}(\mu{\bf r}_{\rm ep}, \mu{\bf r}_{\bar{\rm e}\bar{\rm p}}; \mu R)
\rangle\Big|_{\mu=1}=
\frac{\rm d}{\rm d\mu}E_{\rm BO}(\mu R)\Big|_{\mu=1}. \nonumber \\
\end{eqnarray}
Once again using the fact that $V({\bf r}_{\rm ep},{\bf r}_{\bar{\rm e}\bar{\rm p}},{\bf R})$ is a homogeneous function of order $-1$, one obtains the following form of the virial theorem
\cite{Lowdin:59,Herrmann:67}
\begin{equation}
2\mathcal{T}(R)+\mathcal{V}(R)+R\frac{{\rm d}E_{\rm BO}(R)}{{\rm d}R}=0\,,
\end{equation}
which proves that the expression in the square brackets in (\ref{Eexpan1}) equals zero. Finally $\tilde{E}_{\rm lep}(R)$ reads
\begin{equation}
\tilde{E}_{\rm lep}(R)=E_{\rm BO}(R)+\mathcal{E}_{\rm ad}^{(c)}(R)+\mathcal{O}(m_{\rm p}^{-2})\,.
\label{Eleptfact}
\end{equation}
So the mass-scaled leptonic energy $\tilde{E}(R)$ consists of the Born-Oppenheimer energy $E(R)$, a part of the adiabatic correction $\mathcal{E}_{\rm ad}^{(c)}(R)$ and some other higher order corrections, which should be classified as nonadiabatic contributions. These nonadiabatic corrections can be written in the following explicit form
\begin{eqnarray}
&\tilde{E}_{\rm lep}(R)-E_{\rm BO}(R)-\mathcal{E}_{\rm ad}^{(c)}(R)= \nonumber \\&=
\sum_{n=2}^\infty\left(\frac{-1}{m_{\rm p}}\right)^n\sum_{k=0}^n\left(\frac{n!}{k!}\right)^2\frac{R^k}{(n-k)!}\frac{{\rm d}^k}{{\rm d}R^k}E_{\rm BO}(R)\,.
\end{eqnarray}

\begin{table*}
\caption{Comparison of the non-Born-Oppenheimer correction obtained by the mass-scaling procedure ($\delta E_{\rm lep}(R)$) and the adiabatic correction calculated by Strasburger \cite{Strasburger:04} ($\mathcal{E}_{\rm ad}(R)$) -- in milihartrees.}
{\scriptsize
\begin{tabular}{D{.}{.}{4} D{.}{.}{8} D{.}{.}{8} D{.}{.}{8} p{1mm} D{.}{.}{3} D{.}{.}{9} D{.}{.}{8} D{.}{.}{8}}
\hline
\multicolumn{1}{c}{$R$} & \multicolumn{1}{c}{$\delta E_{\rm lep}(R)$} &
\multicolumn{1}{c}{$\mathcal{E}_{\rm ad}(R)$} &
\multicolumn{1}{c}{$\Delta(R)$} &&
\multicolumn{1}{c}{$R$} & \multicolumn{1}{c}{$\delta E_{\rm lep}(R)$} &
\multicolumn{1}{c}{$\mathcal{E}_{\rm ad}(R)$} &
\multicolumn{1}{c}{$\Delta(R)$}\\
\hline
\hline
0.744 & 0.1441275 &     5934.3     &  -5934.2      &&  2.7   &  0.5395071 &     0.5328086  &     0.0066985 \\
0.746 & 0.1445682 &     6513.1     &  -6513.0      &&  2.8   &  0.5410885 &     0.5338338  &     0.0072547 \\
0.748 & 0.1450162 &     2067.8     &  -2067.7      &&  2.9   &  0.5422054 &     0.5347878  &     0.0074176 \\
0.75  & 0.1454715 &     1469.23    &  -1469.08     &&  3.0   &  0.5429322 &     0.5356682  &     0.0072640 \\
0.8   & 0.1591394 &     40.385     &    -40.226    &&  3.1   &  0.5433773 &     0.5364731  &     0.0069042 \\
0.85  & 0.1767098 &     11.8294    &    -11.6527   &&  3.2   &  0.5436549 &     0.5372086  &     0.0064463 \\
0.9   & 0.1972185 &     5.656181   &    -5.458963  &&  3.3   &  0.5438564 &     0.5378795  &     0.0059769 \\
0.95  & 0.2195822 &     3.36107    &     -3.14149  &&  3.4   &  0.5440330 &     0.5384894  &     0.0055436 \\
1.0   & 0.2427954 &     2.27030    &     -2.02750  &&  3.5   &  0.5441933 &     0.5390447  &     0.0051486 \\
1.1   & 0.2887498 &     1.310081   &    -1.021331  &&  3.6   &  0.5443153 &     0.5395488  &     0.0047665 \\
1.2   & 0.3311600 &     0.9238342  &    -0.5926742 &&  3.8   &  0.5443107 &     0.5404270  &     0.0038837 \\
1.3   & 0.3685480 &     0.7397641  &    -0.3712161 &&  4.0   &  0.5438799 &     0.5411558  &     0.0027241 \\
1.4   & 0.4005566 &     0.6437032  &    -0.2431466 &&  4.2   &  0.5432116 &     0.5417607  &     0.0014509 \\
1.5   & 0.4273892 &     0.5908774  &    -0.1634882 &&  4.5   &  0.5426022 &     0.5424800  &     0.0001222 \\
1.6   & 0.4496878 &     0.5611774  &    -0.1114896 &&  5.0   &  0.5436598 &     0.5433068  &     0.0003530 \\
1.7   & 0.4683043 &     0.5444602  &    -0.0761559 &&  5.5   &  0.5444552 &     0.5438155  &     0.0006397 \\
1.8   & 0.4839570 &     0.5352240  &    -0.0512670 &&  6.0   &  0.5438428 &     0.5441254  &    -0.0002826 \\
1.9   & 0.4970463 &     0.5304100  &    -0.0333637 &&  6.5   &  0.5436756 &     0.5443113  &    -0.0006357 \\
2.0   & 0.5077419 &     0.5282176  &    -0.0204757 &&  7.0   &  0.5440693 &     0.5444247  &    -0.0003554 \\
2.1   & 0.5162030 &     0.5275680  &    -0.0113650 &&  8.0   &  0.5442068 &     0.5445352  &    -0.0003284 \\
2.2   & 0.5227266 &     0.5278091  &    -0.0050825 &&  9.0   &  0.5442654 &     0.5445789  &    -0.0003135 \\
2.3   & 0.5277358 &     0.5285403  &    -0.0008045 &&  10.0  &  0.5442959 &     0.5445979  &    -0.0003020 \\
2.4   & 0.5316630 &     0.5295251  &     0.0021379 &&  12.0  &  0.5443144 &     0.5446109  &    -0.0002965 \\
2.5   & 0.5348346 &     0.5306166  &     0.0042180 &&  15.0  &  0.5443191 &     0.5446156  &    -0.0002965 \\
2.6   & 0.5374283 &     0.5317269  &     0.0057014 &&  20.0  &  0.5443203 &     0.5446168  &    -0.0002965 \\
\hline
\label{tableAD}
\end{tabular}
}
\end{table*}

The adiabatic effects included by the mass-scaling procedure come from the monomer part of the $H'$ operator. The two other parts (given as expectation values of 
$H'_{(a)}$ and $H'_{(b)}$ operators with the leptonic wave function) are not included in the
$\tilde{E}(R)$, however, the $\mathcal{E}_{\rm ad}^{(c)}$ correction is important to correctly reproduce the asymptotical behaviour of the adiabatic potential. For large internuclear separations $R$ the interaction part of the Born-Oppenheimer energy decays as
\begin{equation}
E_{\rm BO}(R)-E_{\rm BO}(\infty)=-C_6R^{-6}-C_8R^{-8}-C_{10}R^{-10}+\dots
\end{equation}
Due to the scaling relation (\ref{scaling2}) the interaction part of the mass-scaled leptonic energy for large $R$ behaves as
\begin{eqnarray}
\tilde{E}_{\rm lep}(R)-\tilde{E}_{\rm lep}(\infty)&=-\tilde{C}_6 R^{-6}-\tilde{C}_8 R^{-8}-\tilde{C}_{10} R^{-10}+\dots \nonumber \\&=
-\frac{C_6}{\mu^5}R^{-6}-\frac{C_8}{\mu^7}R^{-8}-
\frac{C_{10}}{\mu^9}R^{-10}+\cdots
\end{eqnarray}
Expanding $\tilde{C}_6$ in powers of $m_{\rm p}^{-1}$ one gets
\begin{equation}
\tilde{C}_6=C_6+\frac{5}{m_{\rm p}}C_6+\mathcal{O}(m_{\rm p}^{-2})=C_6+\delta C_6^{\rm ad}+
\mathcal{O}(m_{\rm p}^{-2})\,,
\end{equation}
where $\delta C_6^{\rm ad}=5C_6/m_{\rm p}$ is the adiabatic correction to the $C_6$ constant \cite{Dalgarno:56}.

It should also be stressed that after the scaling procedure the leptonic energy curve has a proper nonadiabatic dissociation limit
\begin{equation}
\lim_{R\to\infty}\tilde{E}_{\rm lep}(R)=-\mu \,,
\end{equation}
i.e. twice the nonadiabatic energy of the ground state of the hydrogen atom. This is a simple consequence of using $\tilde{H}_{\rm lep}$ instead of $H_{\rm lep}$ since the former contains the nonadiabatic Hamiltonian for the monomers.

In Table \ref{tableAD} we compared the values of the adiabatic correction $\mathcal{E}_{\rm ad}$ calculated by Strasburger \cite{Strasburger:04} with the correction obtained from the scaling procedure $\delta E_{\rm lep}(R)$ which is the difference between the mass-scaled and the Born-Oppenheimer energy
\begin{equation}
\delta E_{\rm lep}(R) = \tilde{E}_{\rm lep}(R)-E_{\rm BO}(R)\,
\end{equation}
and $\Delta(R)$ is defined as
\begin{equation}
\Delta(R) = \delta E_{\rm lep}(R) - \mathcal{E}_{\rm ad}(R)\,.
\end{equation}

For large internuclear separations $\Delta(R)$ tends to the difference between the nonadiabatic and the adiabatic dissociation threshold (i.e.  $-1 +1/m_{\rm p}$)
\begin{equation}
\lim_{R\to\infty}\Delta(R)=-\mu +1 - \frac{1}{m_{\rm p}}=\frac{1}{m_{\rm p}+1}-\frac{1}{m_{\rm p}}\,.
\end{equation}
This difference comes from the nonadiabatic contributions $\mathcal{O}(m_{\rm p}^{-2})$ which are included in 
(\ref{Eleptfact}).

As it can be seen from Table \ref{tableAD} the scaling procedure reproduces over 95\% of the total adiabatic correction for $R$ larger than 2.0 bohrs. For smaller internuclear distances the interaction part of the adiabatic correction becomes larger and the difference grows. However, in this region the adiabatic approximation starts to fail, and as $R$ gets close to the critical distance $R_{\rm c}$, the adiabatic correction diverges. Since the $\mathcal{E}_{\rm ad}^{(c)}$ correction included in our scaled energy does not diverge for $R\to R_{\rm c}$, one can conclude that the other contributions are responsible for this divergent behaviour of the adiabatic correction near the critical distance. This conclusion seems to be in a good agreement with the reasoning presented by Strasburger. It is shown in \cite{Strasburger:04} that with a rough approximation the expectation value of the $\nabla_{\bf R}$ operator with the leptonic wavefunction is proportional to the mean value of the distance between leptons and nuclei. As $R$ tends to $R_{\rm c}$ the leptonic wavefunction becomes more and more diffused due to the fact that leptons become more weakly bound. In the limit when $R$ reaches the critical value, the leptons are no longer bound to the nuclei and the mean value of the distance between them is infinite. Since the $H'_{(c)}$ operator does not contain a differentiation over the internuclear distance $R$, the $\mathcal{E}_{\rm ad}^{(c)}$  part of the adiabatic correction is not affected by the mechanism described above and is well defined for any $R$. The other two contributions, i.e., $\langle H'_{(a)} \rangle$ and $\langle H'_{(b)} \rangle$, are supposed to diverge as $R$ tends to $R_c$.

\end{document}